\documentclass{emulateapj}
\slugcomment{}
\shorttitle{Cosmic evolution of the M$_{\rm BH}-\sigma$ relation}
\shortauthors{Woo et al.}

\begin{document}

\title{Cosmic Evolution of Black Holes and Spheroids. I: the
M$_{\rm BH}-\sigma$ relation at $z=0.36$}

\author{Jong-Hak Woo\altaffilmark{1}, Tommaso Treu\altaffilmark{1},
Matthew A. Malkan\altaffilmark{2}, Roger D. Blandford\altaffilmark{3}}

\altaffiltext{1}{Department of Physics, University of California,
Santa Barbara, CA 93106-9530; woo@physics.ucsb.edu,
tt@physics.ucsb.edu} \altaffiltext{2}{Department of Physics and
Astronomy, University of California at Los Angeles, CA 90095,
malkan@astro.ucla.edu} \altaffiltext{3}{Kavli Institute for Particle
Astrophysics and Cosmology, Stanford, CA, rdb@slac.stanford.edu}

\begin{abstract}

We test the evolution of the correlation between black hole mass and
bulge velocity dispersion (M$_{\rm BH}-\sigma$), using a carefully
selected sample of 14 Seyfert 1 galaxies at $z=0.36\pm0.01$.  We
measure velocity dispersion from stellar absorption lines around Mgb
(5175\AA) and Fe (5270\AA) using high S/N Keck spectra,
and estimate black hole mass from the H$\beta$ line width and the
optical luminosity at 5100\AA, based on the empirically calibrated
photo-ionization method.  We find a significant offset from the local
relation, in the sense that velocity dispersions were smaller for
given black hole masses at $z=0.36$ than locally. We investigate
various sources of systematic uncertainties and find that those cannot
account for the observed offset. The measured offset is $\Delta \log
M_{\rm BH}=0.62 \pm 0.10 \pm 0.25$, i.e. $\Delta \log \sigma=0.15 \pm
0.03 \pm 0.06$, where the error bars include a random component and an
upper limit to the systematics.  At face value, this result implies a
substantial growth of bulges in the last 4 Gyr, assuming that the
local M$_{\rm BH}-\sigma$ relation is the universal evolutionary
end-point.  Along with two samples of active galaxies with
consistently determined black hole mass and stellar velocity dispersion taken
from the literature, we quantify the observed evolution with the best
fit linear relation, $\Delta \log M_{\rm BH} = (1.66\pm0.43)z +
(0.04\pm0.09)$ with respect to the local relationship of Tremaine et
al. (2002), and $\Delta \log M_{\rm BH} = (1.55\pm0.46)z +
(0.01\pm0.12)$ with respect to that of Ferrarese (2002).  This result
is consistent with the growth of black holes predating the final
growth of bulges at these mass scales ($\langle\sigma\rangle$=170 km
s$^{-1}$).  \end{abstract}


\section{Introduction}

Understanding how and when galaxies form has been a central problem in
cosmology for the past few decades (e.g. White \& Rees 1978, Blumenthal et
al.\ 1984). One of the main recent developments of this field has been
the realization of the key role played by nuclear activity due to
accretion onto supermassive black holes (BHs). Nuclear activity and 
supermassive BHs have been suggested to be relevant for a
variety of phenomena, including regulating star formation at galaxy
scales via energy feedback to solve the `downsizing' problem (Cowie et
al. 1996; Treu et al. 2005a,b; Bundy et al. 2005; Juneau et al. 2005;
De Lucia et al.\ 2006), providing extra energy to solve the `cooling
flow' problem (e.g. Dalla Vecchia et al. 2004, and references
therein), and altering the inner regions of early-type galaxies to solve
the `inner cusp/core' problem (Milosavljevic, et al.\ 2002; Graham
2004; Boylan-Kolchin, Ma \& Quataert 2004).

The best supporting evidences for a strong
connection between galaxy formation and nuclear activity are the tight
empirical scaling relations connecting the mass of the central supermassive
BH with global properties of the host galaxy, such as bulge
luminosity (McLure \& Dunlop 2002; Marconi \& Hunt 2003)
and mass (Magorrian et al. 1998, H\"aring \& Rix 2004),
galaxy-light concentration (Graham et al. 2001), and stellar velocity
dispersion (Gebhardt et al. 2000; Ferrarese \& Merritt 2000). The
latter correlation, the so-called M$_{\rm BH}-\sigma$ relation
has received particular attention for its tightness, which has
been suggested to be consistent with no intrinsic scatter (but see
Novak, Faber \& Dekel 2006 for a statistical analysis of the existing
correlations).

The extension of the known empirical correlations between galaxy
properties at kpc scales (such as the Fundamental Plane,
Dressler et al. 1987; Djorgovski \& Davis 1987) to the pc scales of
the sphere of influence of the central BH indicates that the
formation and evolution of galaxies and supermassive BHs are
connected at a fundamental level.

The variety of scales involved makes this a challenging topic to study
from both a theoretical and an empirical point of view.  From a
theoretical point of view, even the most advanced numerical
cosmological simulations (e.g. Springel et al. 2005) do not
have enough dynamical range to resolve the scales relevant to the
central BH. Thus, theoretical studies generally follow one of
these two approaches (see, e.g., Stiavelli 1998, Miralda-Escud\'e \&
Kollmeier 2005, and references therein for different approaches): i)
studying the scaling relations of remnants of mergers of cosmologically
inspired progenitors, simulated in as much physical detail as
possible (e.g. Robertson et al. 2005); ii) adopting a semi-analytic
approach where the effects of nuclear activity are coupled to dark
matter merging trees via analytic empirical recipes, similar to the
way interstellar medium and stars are treated (e.g. Kauffmann \&
Haenelt 2000; Volonteri et al. 2003; Menci et al. 2003; Vittorini,
Shankar \& Cavaliere 2005; Croton et al. 2005; Cattaneo et al. 2005).

The results of these calculations clearly illustrate the
complexity of the problem. Even imposing the local scaling relations
as the end-point of the coevolutionary models, 
many 
diverse results
can be obtained. For example, in terms of numerical simulations of
mergers, it appears that the scaling laws can be either preserved or
destroyed during a merger, depending on the properties of the
progenitors (e.g. dry or gas rich), and the uncertain fate of the
central BHs (merger with mass conservation, mass-loss through
gravitational radiation, and ejection). When these results are placed into
a cosmological framework for the progenitors, a prediction on the
evolution of the M$_{\rm BH}-\sigma$ relation can be obtained. For
example, Robertson et al.\ (2005) predict that the relation should
evolve in the sense of velocity dispersion increasing with redshift
for a fixed BH mass. Analogously, depending on the details of
the modeling of gas accretion on the central BH and assembly
of the stellar mass, semi-analytic models can predict the evolution of
the scaling relations, with a variety of outcomes. For example,
Kauffmann \& Haehnelt (2000) find the M$_{\rm BH}-\sigma$ relation to
be constant with redshift -- providing a natural explanation for its
tightness-- while Croton (2005) finds that stellar mass (and hence presumably
velocity dispersion)
decreases as a function of redshift for a fixed BH mass.      

From the observers' point of view, three main lines of inquiry are
being pursued to reconstruct the coevolution of spheroids and BHs.
Firstly, improving the determination of the local scaling laws,
e.g. extending the samples in size and dynamic ranges, measuring their
coefficient, scatter and distribution of residuals. The local
relationships are the foundation of the BH-galaxy connection, and
contain important evolutionary clues in the form of `fossil
evidence'. For example, Nipoti et al.\ (2003) find it impossible to
preserve all scaling laws during collisionless mergers, and use this
finding to rule out dry mergers as the only mechanism for spheroid
formation (see also Ciotti \& Van Albada 2001; Kazantsidis et
al. 2005; Treu et al.\ 2006). In addition -- if selection effects are
properly taken into account -- the intrinsic scatter of the
correlation can be used to constrain formation models (Robertson et
al. 2005; Wyithe \& Loeb 2005). Increasing the number of objects with
well-measured BH mass -- through stellar and gas kinematics for
inactive BHs or reverberation mapping for active nuclei-- is
painstakingly difficult with present technology, but nevertheless
important progress is being made (see, e.g., Peterson et al. 2005) and
substantial improvement is expected with the next generation of
telescopes and adaptive optics systems.

The second line of evidence (e.g. Soltan 1982) is provided by the
demographics of galaxies and active galactic nuclei (AGNs). Since the
energy irradiated by nuclear activity is often supposed to be
proportional to the mass accretion rate onto the BH, the luminosity
function of AGN and its evolution measure the build up of BHs,
although the mass accretion rate could be much higher at high
redshift.
Among other results -- e.g. on the consistency of the mass
function of local BHs with that inferred from the accretion history
(Salucci et al.\ 1999; Yu \& Tremaine 2002; Marconi et al. 2004;
Shankar et al.\ 2004)-- this kind of arguments points towards an
anti-hierarchical (`downsizing') growth of BHs, with the more massive
BHs ($>10^8 M_{\odot}$) growing at earlier times ($z\sim1-3$) with
respect to their less massive counterparts (Marconi et al.\ 2004; see
also Small \& Blandford 1992; Tamura, Ohta, \& Ueda 2005; Merloni
2004; Heckman et al.\ 2004). Taking a further step, Merloni, Rudnick
\& di Matteo (2004) combine the BH mass accretion history with the
star formation history of the Universe to find that the growth of
BHs appear to predate that of bulges. In other words, the
bulge-to-BH mass ratio is a decreasing function of redshift,
consistent -- at least qualitatively -- with independent evidence for
significant mass assembly in spheroids since $z \sim 1.5$ (Bell et
al.\ 2005; Treu et al.\ 2005a,b; Bundy et al.\ 2005) and for
similarities between the star formation histories of active and
inactive galaxies (Hunt et al. 1997,1999; Hunt \& Malkan 2004; Woo et
al.\ 2004, 2005). In this scenario, BHs would reach their almost final
mass early-on and then the bulge would grow around them, with the
local scaling-laws as the evolutionary end-point.

The third approach consists of measuring directly the evolution of
scaling laws over cosmic time. The main obstacle to overcome to reach
this goal is the determination of the BH mass at distances
where the sphere of influence (with a radius $\sim GM_{\rm BH}~ \sigma^{-2}$)
cannot be resolved. This approach is
thus limited to active BHs, where the mass of the BHs
can be inferred from the kinematics of the broad emission line region
and its distance from the nucleus (obtained via reverberation mapping
or some empirically calibrated surrogate, see \S~4). In turn, by
restricting the analysis to active BHs, it is more difficult
to measure the properties of the host galaxy, and it is important
to understand any caveats introduced 
when comparing to the local benchmark sample of mainly
non-active galaxies.

In spite of the challenges and the limitations, this approach is the
only one that allows a direct mapping of the coevolution of BHs
and spheroids, and thus is an extremely valuable complement to
the first two described above, lifting many of the degeneracies
introduced by evolution (fossil evidence) or by population averages
(BH and galaxy demographics). For these reasons a few groups
have been attacking the problem using different observational
strategies. Walter et al.\ (2004) use spatially resolved radio
observations to estimate the mass of the host galaxy of a QSO at
z=6.41, with a result almost two orders of magnitude smaller than
predicted by the local BH-spheroid relation 
for the value of BH mass
measured by Willott et al.\ (2003). In contrast, Shields et al.\
(2003) use the width of the [\ion{O}{3}] narrow line as a surrogate for the
stellar velocity dispersion of a sample of bright QSO, finding no evidence for evolution
of the M$_{\rm BH}-\sigma$ relation out to $z\sim3$, although with a large
scatter.

Our group is following a different approach, taking smaller steps in
redshifts and striving to achieve the most accurate possible
measurements of each parameter. As described in our first paper (Treu,
Malkan \& Blandford 2004, hereafter TMB04), we target Seyfert 1
galaxies
to simultaneously measure AGN broad line width and host galaxy stellar
velocity dispersion in specific redshift intervals, where sky emission
and absorption can be avoided (see \S~2.1).  Our pilot study found
tentative evidence of evolution, in the sense that our 7 objects had
smaller velocity dispersion for given BH mass than expected from the
local correlations.

This paper seeks to expand and improve on the pilot study in order to
verify this tantalizing and perhaps surprising finding. The main
improvements over the previous study include: i) a larger dataset (20
objects) with better quality Keck data (S/N up to 100\AA$^{-1}$ as
opposed to $\sim$ 50 \AA$^{-1}$ of the pilot study
and thus smaller uncertainty on the velocity dispersion; ii) higher
completeness to limit possible selection effect (14/20 velocity
dispersions measured as opposed to 7/13); iii) a detailed analysis of
many potential sources of systematic uncertainties, including that due
to the broad nuclear Fe II contamination to the spectral region used
for velocity dispersion measurement; iv) improved estimates of BH
masses based on the extensive work done by Peterson et al.\ (2004) and
Onken et al.\ (2004) to calibrate the scaling relations between
continuum luminosity, H$\beta$ width and BH mass. High resolution
images taken with the Advanced Camera for Surveys (ACS) on board the
Hubble Space Telescope (HST) are also available and will be used in
this paper to interpret the results (a detailed analysis of the
HST-ACS images will be presented in the second paper of this series,
hereafter Paper II).

The paper is organized as follows, In \S~2 we describe sample
selection, observations and data reduction. In \S~3 we describe
stellar velocity dispersion measurements and discuss the viability of
[\ion{O}{3}] as a surrogate for stellar velocity dispersion. In \S~4 we derive the BH
mass and discuss its relation with other AGN properties.  In \S~5 we
present the evolution of the M$_{\rm BH}-\sigma$ relation. Section~6
summarizes a number of tests aimed at investigating possible systematic
uncertainties. In \S~7 we discuss our results and their implications
for the coevolution of spheroids and BHs. Section~8 concludes
with a brief summary.  Throughout the paper we adopt a cosmology with
$\Omega=0.3$, $\Lambda=0.7$, and $H_{0}=70$ km sec$^{-1}$ Mpc$^{-1}$.

\section{Observations and Data Reduction}

\subsection{Sample Selection}

As in our pilot study, the choice of targets is key to the success of
this experiment. On the one hand, the sphere of influence of
supermassive BHs in galaxies at cosmological distances cannot be
resolved even with HST. Therefore, it is necessary to target active
galaxies, where BH mass can be obtained from the integrated
properties of the broad line region -- as described in
Section~\ref{sec:bhm}. On the other hand, in order to 
measure the {\it stellar velocity dispersion} from absorption lines,
we need to target relatively low luminosity AGNs where the fraction of
stellar light in the integrated spectrum is non-negligible. This is
because in distant galaxies, ground-based telescopes cannot separate
the emission of the host galaxy from that of the AGN.

Seyfert 1 galaxies provide the right balance between the two components:
absorption features typical of old stellar populations such as the Mgb triplet ($\sim$ 5175\AA)
and Fe (5270\AA) (Trager et al.\ 1998) are clearly visible in their high
S/N integrated spectra. In order to minimize the
systematic uncertainties related to sky subtraction and atmospheric
absorption corrections, it is convenient to select specific redshift
windows where the relevant emission and absorption lines (H$\beta$,
Mgb, and Fe) fall in clean regions of the Earth's
atmosphere. Accordingly, we selected the ``clean window'' $z = 0.36\pm0.01$,
which corresponds to a look-back time of $\sim 4$ Gyrs.

A first object (MS1558+453; hereafter S99; Stocke et al.\ 1991) was
selected for a pilot study (see TMB04). When the Sloan
Digitized Sky Survey (SDSS) became available, a larger sample of
objects was selected according to the following criteria:
$0.35<z<0.37$, and H$\beta$ equivalent width and gaussian width greater
than 5 \AA\, in the rest frame. All SDSS spectra satisfying these
criteria were visually inspected by two of us (TT and MAM) and objects
showing strong Fe II nuclear emission (the main obstacle to velocity
dispersion measurement) were eliminated from the sample.

The relevant properties of the observed objects are listed in
Table~\ref{T_target}. Although targets were initially selected from
the SDSS DR1 and DR2, all SDSS related quantities listed in this paper
have been updated to reflect all the recalibrations available
from DR4. Throughout this paper we will make limited use of
morphological information derived from an ongoing imaging program with
the Advanced Camera for Surveys on board HST (GO-10216; PI Treu) to
interpret our results. 

\subsection{Observations}      

High S/N ratio spectra of 20 targets were obtained with the LRIS
spectrograph at the Keck-I telescope in five runs between March 2003
and July 2005.  The 900 lines mm$^{-1}$ grating centered at 6700\AA~was 
used most of the time, yielding a pixel scale of 0.85\AA$\times 0\farcs$215 and a
resolution of $\sim$55 km s$^{-1}$. 
For two objects, S28 and S29, we used the 831 lines mm$^{-1}$ grating,
with which $\sim$67 km s$^{-1}$ resolution was obtained.
Observing conditions were
generally mediocre, plagued by cirrus and humidity, except for the
last runs in May 22 2004 and July 2005. Table~\ref{T_journal} shows the journal of
observations and instrumental setups. Typically more than two exposures per
object were obtained to ensure proper cosmic ray removal, with total
exposure times ranging between 600 and 12600 seconds. Taking
advantage of the good observing conditions, the exposure times for the
last observing run were much higher than in previous runs and some
objects with marginal quality spectra were reobserved with longer
exposure times (e.g. S05 was observed for 3 hours in addition to the
0.5 hrs obtained during the September 2003 run).

Internal flat fields were obtained before and after observing each object, so as to correct
the fringing pattern of the red CCD. A set of A0V stars, selected from
the Hipparcos catalog\footnote{URL
http://www.gemini.edu/sciops/instruments/niri/NIRISpecStdSearch.html}
to be within $\lesssim$ 15 degrees from each target, was observed
during the night as secondary flux calibrators and to measure the
B-band atmospheric absorption. Spectrophotometric standards were
observed at the beginning and at the end of each night.

\subsection{Data Reduction}

The standard data reduction processes including bias subtraction,
flat-fielding, spectral extraction, wavelength calibration, and flux
calibration were performed using a series of {\sc IRAF} scripts.
Cosmic rays were removed from each individual exposure using the
Laplacian cosmic ray identification software (van Dokkum 2001).  Sky
emission lines were used for wavelength calibration.

Optimal extraction was used for obtaining one-dimensional spectra with
maximal S/N.  Typical extraction radius was 4--5 pixels, corresponding
to $\sim 1''$ ($\sim 5$ kpc at $z=0.36$).  Individual spectra
were combined with inverse-variance weighting to make the final
spectrum (S/N is in the range of 31 to 111 per 0.85\AA\, pixel for the
combined spectra). The relatively featureless spectra of the A0V
Hipparcus stars were used to correct the atmospheric absorption
(including the prominent B-band) and perform relative flux
calibration.  Internal tests show that this procedure yields the
necessary correction of the B-band absorption to a level of a few
parts in a thousand and relative flux calibration to a few percent
(confirmed also by comparing our spectra to the SDSS spectra).

The final flux-calibrated spectra of all 20 observed AGNs are shown in
Fig.~1. Note the higher S/N ratio of spectra obtained in 2005 with
respect to the older spectra. 

\section{Stellar Velocity Dispersion Measurements}

Stellar velocity dispersions of external galaxies are often measured
from strong absorption lines such as the G-band (4304\AA), 
the Mgb triplet ($\sim$ 5175\AA), Fe (5270\AA), and 
the Ca II triplet ($\sim$ 8550 \AA).  For the host galaxies
of broad line AGNs, the G-band is impractical due to contamination from
H$\gamma$, while the Mgb-Fe and the Ca II triplet regions represent viable
choices, provided spectra of sufficiently high S/N are available
(Nelson \& Whittle 1996; Barth et al. 2003; Woo et al. 2004; 
Greene \& Ho 2005c). In
this study, we focus on the Mgb-Fe region, since the Ca II triplet is
redshifted outside the wavelength range accessible with optical
spectrographs.

Higher S/N than for quiescent galaxies is needed to compensate for the
dilution of stellar features by the AGN continuum, scaling as $1/f_*$
where $f_*$ is the fraction of stellar light in the integrated
spectrum. Systematic errors such as those due to correction for
atmospheric absorption must also be smaller than for quiescent
galaxies by a factor of $f_*$ to account for line dilution. As 
discussed in Section~\ref{ssec:sigma}, the quality of
our spectra is sufficient to recover velocity dispersion with an
uncertainty of 10\% or better.

However, before we can proceed to measure velocity dispersion, we need
to address an important possible source of systematics, contamination
of the AGN power-law continuum by broad nuclear Fe II emission. In our
pilot paper the Fe II contribution was removed using high order
polynomials. In this paper we improve on our Fe II emission subtraction
by using the I Zwicky 1 local Fe II template (Boroson \& Green 1992;
kindly provided by Todd Boroson) as described in the next Section.

\subsection{Fe II Subtraction}

In the case of host galaxies of type 1 AGNs, strong Fe II emission
around 5150-5350\AA~ complicates velocity dispersion measurements
from the Mgb and Fe lines
since the observed continuum shape is very different from that of
template stars.  To reduce potential systematic errors on dispersion
measurements, we decided to subtract broad Fe II emission from the
observed spectra before fitting with template stars (for a consistency check,
we also measured velocity dispersion from spectra without Fe II
emission subtraction. See \S~\ref{ssec:feiicheck}).

We subtracted broad Fe II emission using the I Zw 1 template. First,
following Boroson \& Green (1992), we prepared a set of Fe II templates
with various widths and strengths.  Then, for the purpose of this fit,
we added a power-law spectrum representing locally the featureless AGN
continuum, and a constant representing the stellar
component\footnote{Since the stellar features are much narrower and
often shallower than the nuclear Fe II emission, this is an appropriate
approximation for this purpose}:
\begin{equation}
F(\lambda)  = a \times \lambda^{\alpha} + b \times Fe II(\lambda,v) + g,
\label{eq_Fe II}
\end{equation}
where $\alpha$ is the continuum slope, $v$ and $b$ set width and
strength of Fe II template, and $g$ represents the stellar flux.  For
each set of parameters, the model spectrum was compared with observed
spectra to compute the $\chi^2$.  By $\chi^{2}$ fitting, we derived
the best fit width and strength of broad Fe II emission, and the slope
of AGN continuum as demonstrated in Fig. 2.  Since we are interested
in removing broad Fe II emission to ease the stellar velocity
dispersion measurements, we only used the 5100-5400\AA\, region to fit
the Fe II emission.

The Fe II emission subtracted spectra are compared with the observed
spectra in Fig.~\ref{fig_Fe IIsub}.  We note that there is a range of
Fe II emission strengths (3--24\% in total luminosity at 5100-5500\AA).
Fe II emission is stronger for S01, S03,
S05, S07, S10, S11, S21, S29 while Fe II emission is negligible for
other cases. Stellar absorption features are clearly shown in
many AGN spectra.  However, we can not identify stellar lines for 6
objects, i.e. S02, S03, S10, S21, S27, and S29, most likely due to the
much higher AGN continuum.

\subsection{Stellar Velocity Dispersion Measurements}

\label{ssec:sigma}

Velocity dispersions were obtained by comparing in pixel space the
observed host galaxy spectra with stellar templates broadened with a
Gaussian velocity width ranging between 50-400 km s$^{-1}$. We used as
kinematic templates high resolution spectra of 5 giant stars (spectral types: G8,
G9, K0, K2, and K5) observed with ESI on Keck-II (Sand et al.\ 2004).  Fits
were performed for every template to estimate uncertainties due to
template mismatch, while the best fitting template was used for the
final dispersion measurements.

The minimum $\chi^2$ fit was performed using the Gauss-Hermite Pixel
Fitting software\footnote{Available at
http://www.stsci.edu/~marel/software.html} (van der Marel 1994).  The
merit of fitting in pixel space as opposed to Fourier space is that
typical AGN narrow emission lines (e.g. [Fe VII] 5160\AA, [N I]
5201\AA, and [Fe XIV] 5304\AA; Vanden Berk et al. 2001) 
can be easily masked out (e.g. Barth et
al.\ 2002; TMB04; Woo et al.\ 2004; see
Fig.~\ref{fig_dis}).  Extensive tests, fitting in various spectral
regions including Mgb 5175\AA\, and Fe 5270\AA, were performed to
determine the best spectral regions. As expected, we find that the Fe
line at 5270\AA\, is most useful in the cases when the Mgb triplet is
mostly affected by narrow AGN emission lines (see Greene \& Ho
2005c). Low order polynomials (2-3) were used to model the overall
shape of the continuum.  For 14 out of 20 objects we determined
reliable velocity dispersions, i.e. robust with respect to changes of
template, continuum polynomial order, and fitting region. The final
measurements are listed in Table~\ref{data}. For the subsample in
common with TMB04 the values listed in Table~\ref{data} supersede those in the
pilot study, given the improved data quality and analysis.

\subsection{[\ion{O}{3}] line width vs. stellar velocity dispersion}

Narrow-line clouds have been suggested to be bound predominantly
by the gravitational potential of the host galaxy bulge (Nelson \&
Whittle 1996), providing an opportunity to use the width of
[\ion{O}{3}]$ \lambda$5007 as a readily measurable surrogate for
stellar velocity dispersions. This method has been applied to
higher redshifts and less favorable AGN-to-stellar
light ratios (e.g. Shields et al. 2003).

\begin{figure*}
\plotone{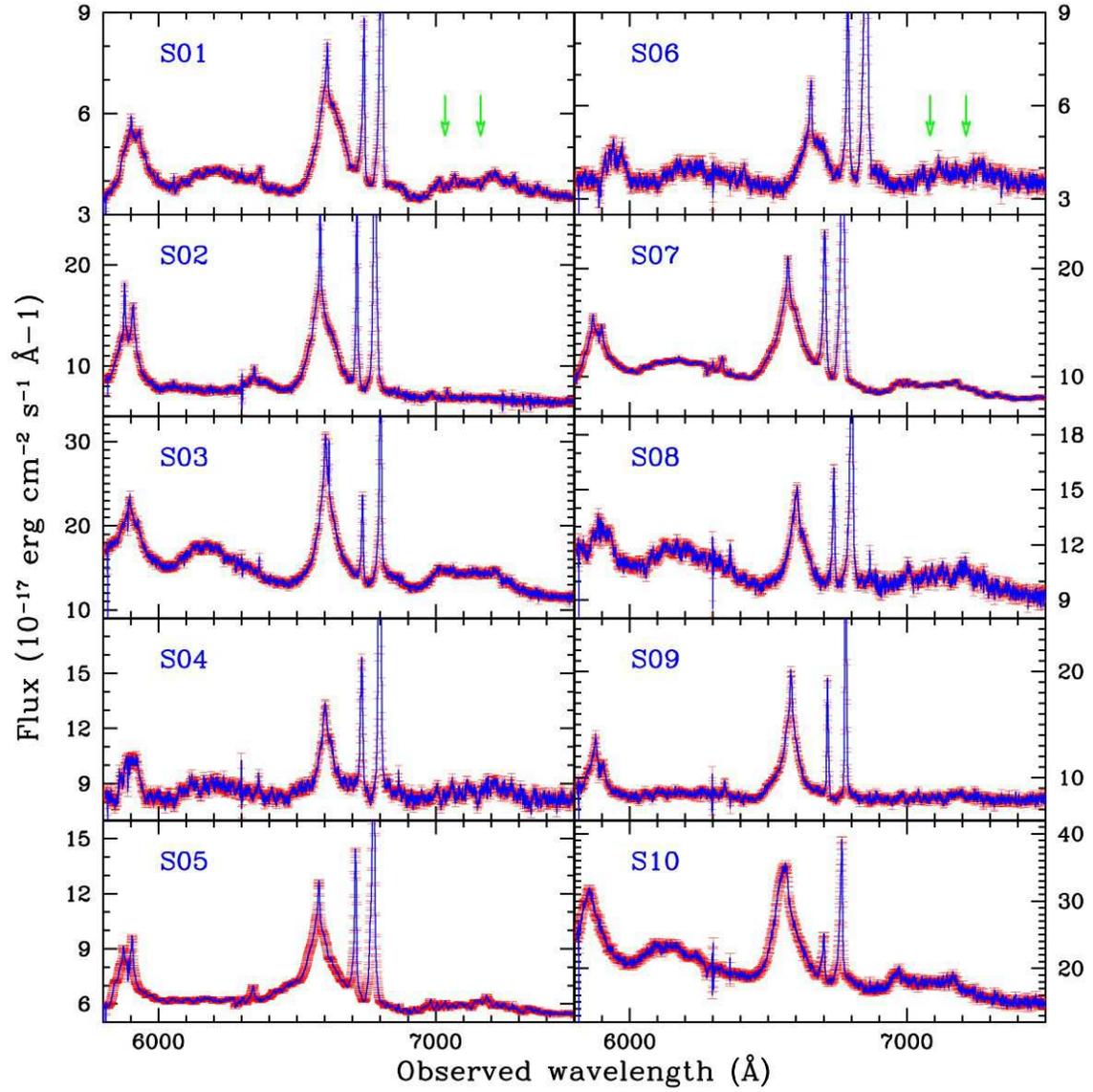}
\caption{Flux calibrated Keck spectra of all 20 Seyfert 1 galaxies. Broad H$\beta$ and
[\ion{O}{3}] lines are clearly present at the center of the wavelength
range. In many cases, stellar absorption features are also visible
redward of [\ion{O}{3}]. The level of noise is indicated by the red error
bars. Green arrows indicate the
location of the stellar features, Mgb (5175 \AA) and Fe (5270 \AA).
}
\label{fig:spectra}
\end{figure*}

The characteristically asymmetric profile of the [\ion{O}{3}] lines
(e.g. Fig.~\ref{fig:spectra}) implies that this assumption is an
approximation and some non-gravitational broadening is present (see Boroson 2005). 
In fact, several studies show that the [\ion{O}{3}] width can trace the
velocity dispersion only in a statistical sense when a large sample is
available (Boroson 2003; Nelson et al.\ 2004; Bonning et al.\ 2005;
Greene \& Ho 2005a; Hicks 2005), and when appropriate care is taken in dealing with
the wings of the line profile.

Here we take advantage of our high quality spectra to revisit the
question of what the best proxy is, i.e. what measure of the
[\ion{O}{3}] line width provides the tightest and least biased
correlation with stellar velocity dispersion.

We measured [\ion{O}{3}] line widths in three different ways, commonly
used in the literature.  First, we fitted [\ion{O}{3}] with a single
gaussian, as typically done with low S/N spectra.  Second, we measured
the FWHM from its definition (e.g. Brotherton 1996; Bonning et al. 2005).  Third, we
removed the blue wing -- present for all 14 objects -- by fitting two
Gaussian components (e.g. Greene \& Ho 2005a).  All these
measurements were done with the {\sc splot} task in {\sc iraf} after
subtracting the continuum.

In Fig~\ref{fig_OIII_sig}, we compare the stellar velocity dispersion
with [\ion{O}{3}] velocities obtained with three different methods,
assuming $\sigma_{OIII} = FWHM/2.35$ when necessary.  As expected,
since all 14 objects show clear asymmetry with a blue wing, a single
gaussian component does not provide a good fit and the proxy generally
overestimates the stellar velocity dispersion by 15\% on average (left
panel), i.e. $\Delta \log \sigma =0.065$.  The simple FWHM does a
better job, reproducing the stellar velocity dispersion with an
average offset of less than 5\% (center panel). In contrast, when the
blue wing is removed, [\ion{O}{3}] line widths are systematically
smaller leading to an underestimate of the stellar velocity dispersion
by almost 20\%.  We find similar scatter for all cases $\sim$ 25\%,
smaller than 0.2 dex found by Nelson \& Whittle (1996; see
also Onken et al.\ 2005), but still a significantly larger uncertainty
than what can be achieved via direct determination (10\%) from stellar
absorption features.

Although based on this test the FWHM appears to be the best
[\ion{O}{3}] based proxy for the stellar velocity dispersion, the
non-gravitational broadening may depend on properties such as
accretion rate (e.g. Greene \& Ho 2005a) and therefore our results 
may 
not apply to {\it all} kinds of AGNs. What appears to be a solid
conclusions is that the intrinsic scatter is larger than typical error
on the stellar velocity dispersion. The results from this section will
be used in \S~\ref{sec:disc} when discussing our results in comparison
with other studies based on the [\ion{O}{3}] proxy.

\section{Black Hole Mass Estimation}
\label{sec:bhm}

The mass of supermassive BHs beyond the local universe cannot
be directly measured from the spatially resolved kinematics because of
the limited spatial resolution of the current instruments.  In the
case of AGNs, the size of the broad-line region, R$_{\rm BLR}$, can be
measured via reverberation mapping (Blandford \& McKee 1982).
Combining the size of the broad line region with its kinematics, as
inferred from the line width, gives the mass of the BH up to a
`shape' factor of order unity which depends on the three-dimensional
geometry and kinematic structure (Peterson 1993; Peterson et al.\
2004). The shape factor has recently been calibrated by Onken et
al. (2004), by requiring active galaxies to obey the same M$_{\rm BH}-\sigma$ relation
as quiescent galaxies.

However, reverberation mapping requires long-term monitoring and it is
extremely expensive to obtain for a large sample of AGNs, especially
at high-redshift. Instead, the so-called `virial' BH mass is
often estimated from a single-epoch spectrum using the empirical
radius-luminosity relation, i.e. the correlation between
R$_{\rm BLR}$ and the optical luminosity at 5100\AA\, 
(Kaspi et al.\ 2000; Kaspi et al.\ 2005). On physical grounds, R$_{\rm
BLR}$ is expected to scale as the square root of the luminosity in
ionizing photons. Although the exact dependence on the optical
luminosity cannot be calculated directly from first principles, it has
been calibrated empirically (Wandel, Peterson \& Malkan 1999; Kaspi et
al.\ 2005). For this reason we will refer to the method used to obtain
BH masses in this paper as empirically-calibrated photo-ionization (ECPI) method.

In this paper we adopt the latest calibrations (Onken et al.\ 2004;
Kaspi et al.\ 2005) of the local reverberation mass shape factor of
the R$_{\rm BLR}$-L$_{\rm 5100}$ relationship, to derive BH
masses:
\begin{equation}
M_{\rm BH} = 2.15 \times 10^{8}{\rm M}_{\odot} \times \left(
{\sigma_{H\beta} \over 3000 {\rm km s}^{-1} } \right)^{2} \left( {\lambda
L_{5100} \over 10^{44} {\rm erg s}^{-1}} \right)^{0.69}~,
\label{eq_BHM} 
\end{equation}
where $\sigma_{H\beta}$ is the second moment of the H$\beta$ line profile
and $L_{5100}$ is the monochromatic luminosity at 5100\AA, in the
rest frame, measured as discussed in the next sections.  Based on
comparisons of reverberation data and single-epoch data (Vestergaard
2002; see also \S~\ref{ssec:hbetatest}), it is estimated that the
intrinsic uncertainty associated with this method is approximately a
factor of 2.5, i.e. 0.4 dex, on the BH mass. This uncertainty
dominates the errors on our input quantities, $\sigma_{{\rm H}\beta}$
and L$_{\rm 5100}$, and we adopt it as our total uncertainty on the
BH mass.        

In closing, we note that the normalization of Onken et al.\ (2004) --
confirmed by Greene \& Ho (2005b) on single-epoch data -- increases
BH mass by 0.26 dex compared to the previous standard shape
factor, (arbitrarily) calculated assuming of isotropic distribution of
broad-line clouds. We emphasize, however, that when measuring the {\it
evolution} of the M$_{\rm BH}-\sigma$ relation, we will use the {\it
same} shape factor for the local and distant sample, so that the
specific choice of the shape factor is irrelevant as long as it is
constant with redshift.

\subsection{H$\beta$ width measurements}
\label{ssec:hbeta}

The width of H$\beta$ is measured in five steps as in our previous
work (TMB04; the procedure is illustrated in
Fig.~6 for all the objects in the sample).  First, we
identify the continuum level on each side of H$\beta$. We typically
used 50\AA\, windows in the range 4690-4780\AA~and 5010-5130\AA~to
measure the the blue and red continuum levels. We then subtracted the
continuum by linear interpolation.  Second, we subtracted
[\ion{O}{3}] $\lambda$4959 by dividing
[\ion{O}{3}] $\lambda$5007 by 3 and blueshifting.  Third, we
rescaled [\ion{O}{3}] $\lambda$5007 and blueshifted it to subtract
the narrow component of H$\beta$.  The flux ratio of
H$\beta$/[\ion{O}{3}] $\lambda$5007 is found to be in the range 6-16\%,
consistent with other studies (e.g. Marziani et al.\ 2003).  Fourth,
we measure the second moment of the line profile for minimum (zero)
and maximum acceptable narrow components. 
When the H$\beta$ profile is very large and clearly goes under [\ion{O}{3}] $\lambda$5007 line,
we obtained this red wing shape by reflecting the blue wing
around the centroid of the broad component.

The statistical noise on the measurement is negligible, given the
quality of the spectra. The three potential sources of systematic
error are removal of the narrow core of H$\beta$, the definition of
the continuum level, and the contribution to the second moment of
broad H$\beta$ from the wavelength region under [\ion{O}{3}]
$\lambda$5007. Varying the intensity of the narrow component of
H$\beta$ changes the second moment by only a few per cent, which is
negligible for our purposes. Changing the wavelength regions defining
the continuum level does not significantly affect the line width
except for the case of S99, where there is a distinct core superposed
on a much broader redshifted component. The low contrast and large
width of the latter component makes its measurement particularly
sensitive to the choice of continuum.  Similar profiles are sometimes
observed in the low states of local Seyfert 1 galaxies such as NGC
3227 and NGC 3516 (Rosenblatt et al. 1994).
If the red continuum were chosen to be at the base between the [OIII]
4959 and 5007 lines, the second moment would be reduced by a factor of 2,
resulting in a smaller BH mass estimate by a factor of 4.
However, as in those local galaxies, it is clear that the H$\beta$ wing
actually does extend to larger positive velocities than 8000 km sec$^{-1}$,
so that the continuum we chose on the red side of 5007 is more
realistic.   
The conclusions of this paper are unchanged
if the narrower line width is adopted for S99.

Three objects, namely, S12, S21 and S23 show a prominently asymmetric
broad-line profile. The extension of the line underneath [\ion{O}{3}]
$\lambda$5007 is therefore unclear, and could result in a small
uncertainty. We measured for these objects line widths with/without
H$\beta$ wing underneath of [\ion{O}{3}] $\lambda$5007, and find less
than 4\% change, suggesting that errors due to red wing of H$\beta$
are negligible.  A final potential source of uncertainty is
contamination from broad Fe II features in the region between 4861 and
5007 \AA\, rest frame. For consistency with the analysis performed on
local calibrators (Kaspi et al.\ 2005), we do not attempt to remove
this component. In section~\ref{ssec:hbetatest} we show that this
source of uncertainty is negligible, by comparing our measurement of
H$\beta$ from single-epoch spectra with that obtained from the rms
spectra of Peterson et al.\ (2004).

\subsection{Luminosity at 5100 \AA}

We directly measured the optical luminosity around 5100\AA\, from the
observed spectra defined as the average flux in the 5070-5130\AA\,
region in the rest frame.  Considering the difficulty of achieving
absolute flux calibration for the Keck spectra -- due to slit losses,
variable seeing and sky transparency-- we tied our spectrophotometry
to the extinction corrected $r'$ band magnitude taken from the
SDSS-DR4 archive.  First, we measured the $r'$ band magnitude from our
observed spectrum using the corresponding response function. Then, we
calculated the offset between Sloan and our measured magnitude, 
to correct the measured flux. 
The mean offset is $0.44 \pm 0.10$.
We did not correct for the stellar contribution to
the luminosity at 5100\AA, to be consistent with reverberation
calibrators. We will discuss the effect of stellar contamination in
\S~\ref{S_L5100error}.

\section{The M$_{\rm BH}-\sigma$ relation at $z=0.36$}

Having measured velocity dispersions and BH masses, in this section we
derive the M$_{\rm BH}-\sigma$ relation of our sample of Seyferts at
$z=0.36$, and compare it with the local samples to infer
evolutionary trends.

Fig.~\ref{fig_msigma} shows our M$_{\rm BH}-\sigma$ in comparison with
the local relationship for quiescent galaxies as measured by Tremaine
et al.\ (2002; solid line; hereafter T02) and by Ferrarese (2002;
dashed line; hereafter F02).  Given the small average difference of
velocity dispersion ($\lesssim$3--4\% for the sample in common) between
T02
and F02,
we neglect aperture effects (see also discussion in Merritt \&
Ferrarese 2001 and in T02). An upper limit to the systematic
uncertainty due to the different apertures adopted in our sample and
those adopted for the local samples is given in \S~6.1.3.

Two things are immediately apparent: 1) all points are above the local
relationship, that is smaller velocity dispersion for a fixed
BH mass; 2) apart from an overall offset and the
relatively narrow range in BH mass, the relationship appears
relatively tight even at this redshift, as we will quantify later in this
section. The small scatter (0.35 dex in BH mass) -- reduced with respect to our previous
measurement (TMB04) -- is encouraging and consistent with a
decrease of the uncertainties due to the improved data quality and
analysis.

\subsection{Redshift evolution}

In order to improve the measurement of the offset from the local relationship and the
amount of scatter, and to assure that we are comparing the same kind of
objects,
we introduce in Fig.~\ref{fig_all} two additional comparison samples,
composed of active galaxies at lower redshift: the 14 reverberation mapped
AGNs with mean redshift of 0.02 from Onken et al.\ (2004), 
and the 15 dwarf Seyfert galaxies with mean redshift of 0.08 from Barth et al.\
(2005).  BH masses of our sample and Barth et al.\ are consistently
estimated using Eq.~\ref{eq_BHM}, which is calibrated on the
reverberation masses available for the Onken et al.\ sample (see
Greene \& Ho 2005b for an additional check on the local calibration of
the ECPI method masses). In other words, a change in the shape factor
will move the three samples vertically by the same amount.

By design, the Onken et al.\ points straddle the local
relationships. The Barth et al.\ points tend to lie preferentially
above the local relationships with an average offset, of which the exact amount depends on the local
slope. The $z=0.36$ points are definitely above the local
relationship. This is illustrated in Fig.~\ref{fig_evolution}, where
we show the offset from the local relationship of T02 as a function of
redshift for all the points in the individual samples.  The average
and rms scatter of each sample is shown as solid red symbols with
error bars.

The offset is clearly detected and appears to increase with
redshift. The best linear fit to the data is shown as a solid line
(for the three samples) and a dashed line (excluding the sample of
dwarf Seyferts from Barth et al. 2005). The best fit linear
relationship for all three samples is $\Delta \log M_{\rm BH} = (1.66\pm0.43)z +
(0.04\pm0.09)$. The rms scatter of the $z=0.36$ sample is 0.35 dex,
similar to that of the Onken et al. sample and to the estimated
uncertainty of the BH masses via ECPI. The average offset of the
$z=0.36$ sample is $0.62\pm0.10$ dex in BH mass, corresponding to
$0.15\pm0.03$ in $\Delta \log \sigma$. A somewhat larger scatter is
found for the Barth et al. sample (0.5 dex), but still remarkably
small, considering that their measurement relies on an extrapolation
of the radius-luminosity relation outside the parameter space of the
reverberation mapped calibrators.

Adopting the local relationship of F02 does not significantly
change the offset of the Onken et al.\ points and marginally
changes the offset of our points ($0.57\pm0.11$ dex) while it
does increase the offset of the Barth et al.\ points from $0.32\pm0.15$ to
$0.45\pm0.15$ dex. This is due to the difference of the local M$_{\rm
BH}$-$\sigma$ relationships for BH masses of order $10^{6}
M_{\odot}$. Including only our points and the Onken et al. points, the
best fit offset with respect to the F02 relation is
$\Delta \log M_{\rm BH}=(1.25\pm0.49)z+0.17\pm0.10$. Even in this
case, we detect evolution in the form of a non-zero slope.  For all
three samples, the linear relationship is $\Delta \log M_{\rm
BH}=(1.55\pm0.46)z+0.01\pm0.12$.

A final important point concerns the morphological type of the host
galaxies. The local samples of T02 and F02 are composed of mainly
elliptical and S0 galaxies, while our sample comprises mostly spiral
galaxies (see Table~\ref{data}). Thus, if -- as discussed in
Section~\ref{sec:disc}-- the M$_{\rm BH}-\sigma$ relation depends on
morphological type, comparing our sample to the local quiescent sample
could introduce a selection bias. However, since the comparison sample
of Onken et al.\ is quite similar to ours in terms of distribution of
host galaxy morphological types, this does not appear to be a
problem. Within the limited size of our sample, we can explore a
possible dependency of the M$_{\rm BH}-\sigma$ relation on
morphological type by considering the offset for two distinct
subgroups, the bulge-dominated (i.e. ellipticals, lenticulars, and Sa;
red solid squares in Fig.~\ref{fig_msigma}) and the disk-dominated
(Sb, Sc; solid blue triangles). For the 4 bulge-dominated systems we
find an average offset of $0.42 \pm 0.16$ dex in BH mass, indicating
that the offset of bulge-dominated systems is still significant.  In
the case of 9 disk-dominated systems we find an offset of $0.75 \pm
0.10$ dex. Within the small number statistics of the subsamples, this
suggests that bulge-dominated systems are closer to the local 
M$_{\rm BH}-\sigma$ relation, consistent with a scenario that the local relation is the
end-point of galaxy evolution as we will discuss in \S~7.

\section{Testing systematic uncertainties}

Before discussing the interpretation of our results in
Section~\ref{sec:disc}, we 
need to understand systematic errors.
In this section we list a number
of potential sources of error and estimate as accurately as possible
the associated uncertainties. We pay particular attention to those
effects that could simulate artificially our observed evolution,
i.e. those that could lead us to overestimate BH mass or to underestimate
the velocity dispersion. To estimate the relevance of the uncertainties, it is useful
to keep in mind that the observed offset is 0.62$\pm$0.10 in
$\Delta \log$ M$_{\rm BH}$ or equivalently 0.15$\pm$0.03 in $\Delta \log \sigma$.
 
\subsection{Are velocity dispersions underestimated?}

\subsubsection{Template mismatch}

Stellar velocity dispersions of early-type galaxies are typically
measured by comparison with broadened spectra of Galactic giant stars,
which are believed to dominate their integrated light. However,
individual stars cannot reproduce perfectly the integrated stellar
populations, and template mismatch has long been known as a potential
source of error (e.g. Rix \& White 1992). To estimate this effect we
use a set of templates covering a range of stellar types and find that
velocity dispersion is stable, varying by less than the estimated
error from one template to another.

Another potential source of error is the mismatch in abundance ratio
between galactic templates and those in external galaxies. For
example, Barth et al.\ (2003) and Woo et al.\ (2004, 2005) found that
the Mgb triplet is much stronger in host galaxies of BL
Lac objects -- which are generally massive galaxies ($ >
10^{11} M_{\odot}$) -- due to the $\alpha$ element enhancements, and
that trying to fit simultaneously Mgb and Fe could introduce systematic
uncertainties (see also Greene \& Ho 2005c).  Only one of our spectra,
S24, shows significant Mgb mismatch, consistent with the overall
similarity between the host galaxies and the Milky Way in 
terms of morphology (see Table~\ref{data}). In
the case of S24, we excluded the Mgb region from the fit (see
Fig.~\ref{fig_dis}). For robust measurements, we tested potential systematics
due to subtle mismatches on the other spectra by repeating the fits
without Mgb and found no significant changes.

We conclude that template mismatch does not introduce significant
systematic errors.

\subsubsection{Continuum fit}
\label{ssec:feiicheck}

The presence of an active nucleus manifests itself in the observed
spectra as two main features: 1) a featureless continuum, which
dilutes stellar absorption lines; 2) various broad and narrow emission
lines.  For our sample and in our wavelength range, the AGN features
that are relevant to the stellar velocity dispersion measurement are
broad Fe II emission at 5100-5400 \AA\, and narrow emission lines in
the vicinity of the Mgb triplet. The narrow lines can be dealt with by
masking out during the kinematic fit. The robustness of velocity
dispersion with respect to masking out the whole Mgb triplet region
shows that the narrow lines do not introduce a substantial source of
error.

To check the effect of subtracting Fe II, we repeated the measurement
of velocity dispersion, using spectra without Fe II subtraction.  In
this case, we normalized the spectra with a high (5-7) order
polynomial fit, before comparing with broadened template stars.  The
two measurements are compared in Fig.~\ref{fig_com_sig}.  We find the
largest difference ($\sim 20\%$) for S07, which has the strongest Fe II
emission. Overall, the agreement is quite good, the two different
measurements agree with a mean offset of 0.01 dex and a rms scatter of
0.04 dex, much smaller than the offset 0.15 dex in log $\sigma$.  The
agreement between the two results suggests that the stellar velocity
dispersions are not underestimated due to systematic effects related
to the Fe II subtraction.

\subsubsection{Aperture correction}

\label{ssec_aperture}
Our adopted extraction window ($\sim 1''\times2'' = \sim5\times10$
kpc$^2$) is of order of the expected effective radius ($r_{e}$) of the
bulges in our sample. For comparison, the average and median of the
$r_{e}$ for the sample of early-type galaxies and early-spirals
analyzed by Treu et al.\ (2005) -- with velocity dispersion in the
same range as our sample -- are 3.8 and 2.7 kpc, respectively. We
adopt these numbers as our fiducial estimate of the $r_{e}$ to
quantify whether the larger aperture with respect to those in the
local studies could introduce a systematic bias in our measured
velocity dispersions.

For the local sample of quiescent galaxies, Gebhardt et al.\ (2000,
also Tremaine et al. 2002) used an aperture close to $r_{e}$, while
Ferrarese \& Merritt (2000, also Ferrarese 2002) corrected velocity
dispersions to a smaller aperture, 1/8 of $r_{e}$.  However, the
average difference in luminosity-weighted velocity dispersions of the
same sample between Tremaine et al. (2002) and Ferrarese (2002) is
only a few per cent since most of the light comes from the central
part of the bulges (Merritt \& Ferrarese 2001; Tremaine et al. 2002).
Similarly, when comparing our sample with the local samples, we expect
small aperture effects on our luminosity-weighted velocity
dispersions. We note that our kinematics is measured from optimally
extracted spectra with inverse variance weighting and thus more
weighted towards the center than a simple luminosity weighted average.

A conservative upper limit to uncertainty due to the aperture effects
can be derived as follows.  For early-type galaxies, the standard
correction proposed by J{\o}rgensen, Franx \& Kj{\ae}rgaard (1995)
implies insignificant correction factors of 1.0\% and 2.5\% (for the
two estimates of $r_{e}$) from our aperture to the velocity
dispersions measured within $r_{e}$ as in the Gebhardt et al.\
(2000) local sample.  The correction for the smaller aperture adopted
by Ferrarese (2002) is 9.9\% and 11.1\%. However, the effect is
probably smaller when considering seeing effects, luminosity and
optimal extraction. For example, Padmanabhan et al. (2004) constructed
velocity dispersion profiles based on SDSS fibers corresponding to
different fractions of an effective radius, and found them to be flat.
As for spirals, recent measurements of the aperture correction factor
indicate that the effect is indeed negligible (Pizzella et al.\ 2004).
We conclude that aperture effects ($\sim 10\%$ at most, i.e. 0.04 dex
in $\sigma$) are significantly smaller than what would be needed (0.15
dex in $\sigma$) to bring our points at $z=0.36$ in
agreement with the local M$_{\rm BH}-\sigma$ relation.

\subsubsection{Host galaxy morphology and inclination}

The final source of potential systematic error for velocity
dispersion that we consider is contamination by kinematically ``cold'' (i.e. with
velocity dispersion smaller than that typical of the bulge) stars in
the disk. Mgb and Fe lines arise from relatively old stellar populations 
(e.g. Trager et al.\ 2000). Therefore, the contribution from the
younger stars in the disks to these lines should be small with
respect to the contribution to the optical continuum.

Nevertheless, it is conceivable that this contamination could play a
role, especially for the most disk-dominated objects. A precision
measurement of this contribution will have to wait for high spatial
resolution integral field spectroscopy, however we can get a sense of
this effect by exploiting the available HST images. In fact,
inclination plays a major role in determining the contribution of
rotation to the unresolved line width. For a face on-disk, rotation
does not contribute to the line width and therefore disk contamination
is expected to bias towards lower stellar velocity dispersion of the bulge.
For an edge-on disk, the unresolved velocity profile of the
disk has width comparable to the stellar velocity dispersion of the
bulge because they are contrasting the same potential\footnote{For
example, modeling the disk as an exponential (Freeman 1970) and the
bulge as an $r^{1/4}$ (de Vaucouleurs 1948) profile, for an aperture
equal to twice the effective radius and to twice the exponential
length, the ratio between the second moment of the line of sight
velocity and the velocity dispersion is $\sim0.86$.}
and therefore we could expect a smaller effect, or even an increase in
velocity dispersion. For intermediate inclination objects, or for
objects with no evidence of a disk -- hereafter the
intermediate/undefined sample -- we can assume that disk contamination
should introduce no bias in the velocity dispersion.

Fig.~\ref{fig_msigma_inc} shows the M$_{\rm BH}-\sigma$ relation by
color-coded inclination from our analysis on HST images
(Paper II).
Face-on spiral galaxies are shown as blue
triangles, edge-on spirals are shown as open circles, the
intermediate and /undefined sample is shown as solid circles. 
A galaxy, for which HST image is not available, is shown as a black square.

As expected, for a given BH mass, face-on spiral galaxies have lower
velocity dispersion than edge-on and intermediate/undefined galaxies.
Thus, it is possible that velocity dispersion of face-on galaxies are
affected by cool stars with low velocity. If we exclude face on
galaxies, the offset is found to be $0.45\pm0.10$ in BH mass ($0.92
\pm 0.10$ and $0.43 \pm 0.12$, respectively for face-on and
intermediate / undefined galaxies), 0.17 dex smaller than that of the
total sample.  It is remarkable that even the edge-on sample shows an
offset in the same direction as the total sample. The
intermediate/undefined sample in particular should be the cleanest
with respect to disk contamination and still shows an offset from the
local relationship, although of course the error bar is larger due to
the much reduced sample size.

We conclude that disk contamination could reduce the
offset by 0.17 dex, but not eliminate it.

\subsection{Are BH masses overestimated?}

\subsubsection{Is H$\beta$ width overestimated?}

\label{ssec:hbetatest}

We estimate BH masses using the width of H$\beta$ line and
optical luminosity at 5100\AA\, from single-epoch spectra, rather
than the RMS spectra used in the derivation of the radius-luminosity
relation for the reverberation sample (Peterson et al.\ 2004;
Kaspi et al.\ 2005).  Thus, we have to test whether our H$\beta$ width
measurement from single-epoch spectra is consistent with that from the RMS
spectra.

In order to perform this test, we collected spectra for as many as
possible of the objects in the reverberation sample of Onken et
al.\ 2004 (our local calibrators). Specifically, we obtained
single-epoch spectra for 6 Seyfert galaxies from International AGN
Watch website\footnote{URL
http://www-astronomy.mps.ohio-state.edu/$\sim$agnwatch}.  Single epoch
spectra for 3 additional Seyfert galaxies were kindly provided by
Aaron Barth, while mean spectra for 5 Seyfert galaxies were kindly
provided by Bradley Peterson.

Fig.~\ref{fig_com_rev} compares for the 14 objects reverberation
masses (Peterson et al. 2004; Onken et al.\ 2004) with our own BH mass
estimates.  To simplify the test, we adopted $L_{5100}$ from Peterson
et al.\ (2004), so that any discrepancy is only due to the difference
between H$\beta$ width from the RMS spectra and that from single-epoch
(for 9 objects) or mean (for 5 objects) spectra. The reverberation masses
are only slightly higher with a mean difference of
$0.09 \pm 0.07$ dex ($0.05 \pm 0.11$ and $0.16 \pm 0.03$, respectively
for the single-epoch sample and the mean spectra sample), suggesting our 
H$\beta$ measurements are consistent with reverberation results.  
This scatter and offset also provide an upper limit to possible systematics
due to minor differences between our implementation of the algorithm
for measuring the second moment of H$\beta$ line width and that 
used by Peterson et al.\ (2004), including
possible residual Fe II emission underneath the broad H$\beta$.  We
conclude that we do not find any evidence for systematic
overestimation of H$\beta$ width.

\subsubsection{$L_{5100}$ overestimated?}

\label{S_L5100error}    
A potential concern is that L$_{5100}$ could be overestimated if the
host galaxy contribution is not negligible. However, the stellar
component is not removed for the local calibration of the R$_{\rm
BLR}$-L$_{\rm 5100}$ relation based on the reverberation mapped sample
(Kaspi et al. 2005). Therefore, for consistency with their procedure
we do not attempt such a removal. Future work on the analysis of HST
images for our sample (Paper II) and for the
local reverberation mapping sample (GO-10516, PI Peterson) will improve
on this point. 

Overall, given the slow dependence of the BH mass on L$_{\rm
5100}$, we expect any differences to be small. For example, if we
suffered from {\it double} the amount of contamination by stellar
light than that for the sample of local calibrators, the BH
mass would be reduced by 0.1 dex, significantly smaller than the
measured offset of 0.62 dex. Similarly, AGN variability at the
expected level of $\lesssim$20-30\%~(i.e $\lesssim$0.05-0.08 dex) is a
negligible effect (Webb \& Malkan 2000) 
and its effect is already included in the overall
estimate of the uncertainty of the BH mass from single-epoch
data.

\subsubsection{BH mass `shape' factor}

As mentioned previously, the geometry and kinematics of the broad-line
clouds -- which connect the observed broad line profile to the
BH mass as M$_{\rm BH} = f \times (\sigma_{H\beta}^{2}~ R_{\rm BLR} /G)$, 
where R$_{\rm BLR}$ is the distance to the broad-line clouds,
$\sigma_{H\beta}$ is the second moment of the H$\beta$ line profile,
and f is the shape factor
 -- are not well constrained. A common practice in the
literature is to adopt the shape factor expected for an isotropic
distribution (i.e. f=3). Here, we adopt the recent Onken et al.\ (2004) shape
factor, f=5.5, calibrated to fit the local M$_{\rm BH}-\sigma$ relation
using 14 Seyfert galaxies with measured stellar velocity dispersion
and reverberation mass (see also Greene \& Ho 2005b).

Although we adopt this new calibration of the shape factor, the
evolution of the M$_{\rm BH}-\sigma$ relation is {\it independent} of its
numeric value -- as long as it is constant with redshift -- since the
relative offsets among three redshift points in
Fig.~\ref{fig_evolution} are effectively BH mass ratios. If
the assumption of constant shape factor is dropped, many different
interpretations of our result become possible, including that the
M$_{\rm BH}-\sigma$ relation is constant but the shape factor evolves
by a factor of four between $z=0.36$ and today.

Unfortunately, direct measurement of BH masses around active
galaxies at high-redshift are unlikely to become available in the near
future, and therefore the shape factor cannot be directly determined
for the moment.  However, we note that the local Onken et al. sample is
similar to our own in terms of BH mass, orientation (i.e. both
are Seyfert 1s), and accretion rate (Woo \& Urry 2002; see below). 
Therefore if these are the main parameters
driving the geometry, their shape factor should be the same, at least
on average.  For this reason, we will continue in our discussion
assuming that the shape factor is constant,
but the reader is cautioned to keep this caveat in mind.

\subsection{Are these BHs growing rapidly?}

A final check concerns the mass accretion rate of the
BHs. In fact, if the BHs were rapidly growing, this
would need to be taken into account when comparing with the local
relationship. For example, if the local relationship is the final
destiny of our sample, velocity dispersions would need to grow more
than 40\% in the next 4 Gyrs to compensate for the increased mass.

The Eddington ratios are shown in Fig.~\ref{fig_LEDD}, with 
the the Eddington Luminosity, defined as $L_{EDD} = 1.25 \times 10^{38}$ 
M$_{\rm BH}/M_{\odot} {\rm erg}$ $s^{-1}$. 
As usual, 
bolometric luminosity is obtained as 9$\times$L$_{\rm 5100}$ (Peterson
et al. 2004; adopting the larger bolometric correction suggested by Marconi et
al. 2004 would not change our conclusions.)
It is clear that our Seyferts
typically have less than 10\% of Eddington luminosity. Their
mass accretion rates are in the range 0.3--0.7 $M_{\odot} $yr$^{-1}$
assuming a standard 10\% efficiency. For a typical AGN life
time of $<$ 0.1 Gyr (Martini 2004 and references therein), these BHs can 
grow at most by a factor of $\sim 2$, i.e. 0.3 dex. 
This amount of growth would strengthen
the case for evolution in the velocity dispersion, implying that
velocity dispersions would need to increase by $\sim$70 \% in the next
4 Gyrs to conform to the local relationship.

\section{Discussion}
\label{sec:disc}

In this section we interprets our results in terms of the co-evolution
of spheroids and BHs. We consider three possible explanations for our
results, ranked from the most prosaic to the most radical.

1) Systematic Errors. 
So far as the first explanation is concerned, 
in the previous section
we discussed in detail a number of possible systematic uncertainties,
and we could not find any single component large enough to bring our
points in agreement with the Onken et al. points. It is possible that
one or more of these sources of uncertainty are combining to enhance
our measured evolution, but we consider it unlikely that the entire
effect is due to a conspiracy of these sources of error (summing all
the possible contributions listed in \S~6, we estimate that the
total systematic uncertainty is 0.25 dex in BH mass).

Another option is that the ECPI BH masses are less accurate than expected,
and possibly biased, depending on some still unknown
additional parameters other than the H$\beta$ width and L$_{\rm
5100}$. The tightness of our measured relation and other independent
tests (e.g. Greene \& Ho 2005b) are quite encouraging in this respect,
but not yet conclusive. While we will continue our effort to increase
the quality of our measurement in the distant universe, it is clear
that much work remains to be done in the local universe.  The sample
of reverberation mapped AGNs used as local calibrators is still
remarkably small, and covers a very limited range in fundamental
parameters, such as BH mass, eddington ratio, orientation, etc.

2) Selection Effects. As the second explanation, selection effects also range from the
trivial to the profound. Signal-to-noise ratio requirements introduce
a minimum threshold in luminosity. Therefore, if there were a
distribution of luminosities for any given BH mass, we would tend to
bias towards the largest masses via Eq.~2. We made some progress
in this respect, increasing our completeness from 7/13 (54\%) to 14/20
(70\%) between our pilot study and the present one. We also have a
selection effect against strong Fe II emission, since that complicates
stellar velocity dispersion measurements.  It has been speculated that
Fe II correlates with high accretion rate (Boroson \& Green 1992), so
excluding Fe II emitters could bias us against fastly growing and hence possibly
smaller BHs. For the
same reason, local studies tend to use the near infrared Ca II triplet
for stellar kinematics. Systematic differences between Mgb + Fe and Ca
II triplet kinematics could simulate an evolutionary trend, although
they are unlikely to be as large as 0.15 dex (see Barth et al. 2003;
Greene \& Ho 2005c). We plan to increase our completeness and compare
Mgb + Fe and Ca II triplet kinematics via optical and infrared
spectroscopy in the near future.

At a more profound level, the local fundamental benchmark -- the
M$_{\rm BH}$-$\sigma$ relation -- is based on only $\sim 30$ quiescent
objects (e.g. Tremaine et al.\ 2002), most of
them elliptical and lenticular galaxies -- i.e. collisionless and
dynamically relaxed systems, except for 4 spirals (MW, M31, NGC1068, and NGC4258).
Selection effects could be at work to artificially tighten the
relationship. It is plausible to imagine a scenario where the M$_{\rm
BH}$-$\sigma$ relation is an evolutionary endpoint, rather than a time
invariant property of all galaxies. For example, it is possible that
only after a sufficient long time the amount of baryons that has been
converted into stars -- which mergers have relocated from the disk to
the bulge -- and the amount of baryons accreted by the central BH,
will be constant fractions of the total amount available, and
thus proportional to each other. However, the road to the M$_{\rm
BH}$-$\sigma$ relation could be full of traumatic events such as gas-rich
mergers, and during their life time galaxies could have roamed through the
M$_{\rm BH}$-$\sigma$ plane (e.g. Kazantzidis et al.\ 2005). If we
only measured the relation for well-behaved early-type galaxies --
where stellar orbits are easy to measure and to model -- we could
artificially select the objects in their tight final correlation. If
points on the M$_{\rm BH}$-$\sigma$ diagram were selected by their
BH mass -- as opposed to by the mass of the spheroid -- we
could imagine to find objects that lie above the end-point relation,
if they existed.  Since our sample is effectively selected for having
a massive BH, a selection effect of this sort could provide an
explanation of our findings that is still in some sense
``evolutionary''. In this respect, once again, the local samples of
AGNs play a crucial role, being selected for their BH rather
than for their bulge. By comparing our objects with the AGNs selected
sample of Onken et al.\ (2004) -- instead of the Tremaine et al.\
(2002) quiescent sample -- we are minimizing this sort of selection
effect. Ongoing work to extend the sample of AGNs with reverberation
mapping BH masses (Peterson, 2005, private communication) and
to obtain direct mass measurements for some of the same objects 
(Hicks 2005) is extremely important to improve the
determination of the local M$_{\rm BH}$-$\sigma$ relation and its
scatter. Recent results could indicate that the relationship could
have more scatter than imagined, especially at smaller masses
(Peterson et al.\ 2005).

3) Cosmic Evolution. This is probably the most
surprising at first, given that the last 4 Gyrs are in many respects a
rather quiet phase of the life of the universe (consider, e.g., the
sharp decline of the cosmic star formation rate since $z\sim1$, Lilly et
al.\ 1995). However, evolution has been observed in the galaxy
population, at least at masses comparable to those of the galaxies in
our sample. In the velocity dispersion range covered by our sample,
significant evolution and star formation is seen even in early-type
galaxies (Treu et al.\ 2005a,b; van der Wel et al.\ 2005; di Serego
Alighieri et al.\ 2005). There is evidence that the morphological mix
changed significantly in last $\sim$5 Gyr, both in the field
(Bundy et al.\ 2005) and in clusters (Dressler et al. 1997). As far as
AGN hosts are concerned, recent studies show that the hosts of the
most massive BHs have relatively old stellar populations,
perhaps as old as those of massive quiescent ellipticals (Woo et al.\
2004, 2005), but little is known for smaller mass systems.  If
evolution were found to be much more recent for smaller mass system,
this could be an indication of `downsizing' (Cowie et al.\ 1996; Treu et
al. 2005a) for AGN-hosts as well. General arguments have been given
suggesting that the global accretion and star-formation history of the Universe and
the growth of bulges follow the growth of
supermassive BHs (Merloni et al. 2004).

More specifically, the real test of our
result is provided by independent measures of the M$_{\rm
BH}$-$\sigma$ relation. Shields et al.\ (2003) combines an
[\ion{O}{3}]-based estimate of the velocity dispersion with ECPI
estimates of the BH mass to find a result consistent with no evolution
out to $z\sim3$. However, the scatter of their measurement
(i.e. $\sim$ 0.5 dex in the offset from the M$_{BH}-\sigma$ relation)
could soften the evolutionary constraints imposed on the relation.  Furthermore, as
discussed in Section~3.3, for high(low)-luminosity quasars of Shields
et al.\ sample, stellar velocity dispersion could be
over(under)estimated since [\ion{O}{3}] line widths are systematically
larger(smaller) than stellar velocity dispersion with an
increasing(decreasing) function of Eddington ratio (Greene \& Ho
2005a), possibly diluting any signature of evolution.  Furthermore,
the mass and redshift range is completely different from our own,
making it even harder to compare the two measurements.  Walter et al.\
(2004) combines CO-based velocity dispersion with an ECPI based mass
estimate for a system at $z\sim6$, and finds evolution in the same
sense as we do, although certainly this study is more comparable to
that of Shields et al.\ (2003) in terms of mass and redshift range.
Another test is provided by studies of the relationship between BH
mass and host galaxy luminosity or stellar mass for distant galaxies,
which also find evidence for AGNs with smaller bulges than expected
(Magain et al.\ 2005; Peng et al.\ 2005).  

One rationalization of our observations is that most galaxies are formed
as a blue, star-forming, disk or irregular state and that
this phase is terminated by a major merger. 
The consequences of this merger are threefold.
Firstly, there will be an onset of nuclear activity as gas,
either present in the major partner or contributed by the minor partner,
is driven into the accretion disk orbiting the major BH.
This creates the Type I Seyfert galaxies in the state,
as we observe them in our sample, where the mass supply rate is significantly
lower than the Eddington rate.

Secondly, the mass supply builds up to exceed the Eddington rate and
this leads to a powerful, disk-driven outflow which expels cool,
interstellar gas (Silk \& Rees 1998, Blandford 1999), and brings about
the transition from the blue sequence to the red sequence, which has
been postulated by Hopkins et al. (2006) to occur at a characteristic
mass which increases with redshift.  In the model of Hopkins et al.,
the estimated transition stellar mass at $z=0.36$ is $\sim 3 \times
10^{10} M_{\odot}$.  This is consistent within the stellar luminosity
of the galaxies in our sample (we will revisit this point in Paper
II).  Normally, dry mergers of spheroids do not change the velocity
dispersion (Nipoti et al.\ 2003 and references therein; but see
Boylan-Kolchin et al.\ 2006 for more general results). However, it is
our conjecture that the combination of gas dynamical interaction
between interstellar molecular clouds and an outflowing wind, as a
short-lived dynamical phase, will lead to some dissipation within the
stars and a consequent increase in the velocity dispersion. In
addition, the direct transfer of stellar mass from the disk component
of the merging galaxies to the spheroids could further boost the final
bulge mass (Croton 2006) and hence its velocity dispersion.

The third phase may occur considerably later and involves three body
interactions of three or more merging BHs in the center of the
combined galaxy.  This can lead to the ejection of some of the BH mass
(e.g. Favata et al. 2004, Blanchet et al. 2005, Haehnelt, Davis \&
Rees 2006, Hoffmann \& Loeb 2006), further lowering the M$_{\rm
BH}-\sigma$ relation.

In conclusion, it appears that -- at least qualitatively --
dissipational mergers must play an important role in the later stages
of assembly of spheroidals (see also Nipoti et al.\ 2003; Kazantzidis
et al. 2005; Roberston et al.\ 2005; Treu et al.\ 2006; Koopmans et
al.\ 2006) if evolution is the correct interpretation of our
observations.

\section{Summary}

We summarize our results as follows:

1) We test the evolution of the M$_{\rm BH}-\sigma$ relation,
by measuring stellar velocity dispersions with high S/N spectra for a
sample of 14 Seyfert 1 galaxies at $z=0.36\pm0.01$.  BH masses are
estimated using the H$\beta$ line width and the optical luminosity at
5100\AA, based on the empirically-calibrated photo-ionization method.

2) We find a significant offset, $\Delta \log M_{\rm BH} =
0.62\pm0.10$ ($\Delta \log \sigma = 0.15\pm 0.03$) from the local
relation of Tremaine et al. (2002), and $\Delta \log M_{\rm BH} =
0.57\pm0.11$ ($\Delta \log \sigma = 0.14\pm 0.03$) from that of
Ferrarese (2002), in the sense that velocity dispersions were smaller
for given BH masses at this redshift.

3) We investigate various sources of systematic errors, and
find that those cannot account for the observed offset.  Combining
systematic errors of aperture correction ($<$0.15 dex in M$_{\rm
BH}$), contamination from cold disk kinematics (0.17 dex), and stellar
contamination to the optical luminosity at 5100\AA~(0.1 dex), we
estimate an upper limit to the systematic uncertainty of 0.25 dex in
BH mass (0.06 in $\Delta \log \sigma$).

4) Along with two samples of AGNs at lower redshifts, we
quantify the observed evolution with the best fit linear relation,
$\Delta \log M_{\rm BH} = (1.66\pm0.43)z + (0.04\pm0.09)$ with respect
to Tremaine et al. (2002), and $\Delta \log M_{\rm BH} =
(1.55\pm0.46)z + (0.01\pm0.12)$ with respect to Ferrarese (2002),
consistent with the scenario where the BH growth predates bulge assembly at these
mass scales ($\sigma$=120--220 km s$^{-1}$).

5) We compare three measurements of the [\ion{O}{3}]
$\lambda$5007 line widths with the stellar velocity dispersions and find a scatter
of $\sim$25\% (i.e 0.1 dex in $\sigma$, or 0.4 dex in M$_{\rm BH}$),
and systematic offsets depending on the line fitting methods.
These uncertainties must
be accounted for when studying the evolution of the M$_{\rm
BH}-\sigma$ relation, especially with a small sample.

\medskip

In \S~7 we discussed three possible explanations for the observed
offset: 1) systematic errors; 2) selection effects; 3) cosmic
evolution. Systematic errors are unlikely to account for the offset
($\sim$0.60 dex), which is significantly larger than the overall
systematic uncertainty (0.25 dex).  Selection effects could be present
both in our sample (selected against low luminosity AGNs and thus
small BH masses), and in the local sample (favoring more evolved
systems), possibly resulting in the observed evolution of the M$_{\rm
BH}-\sigma$ relation. In order to minimize selection effects, we
compared our z=0.36 galaxies to the sample of local AGNs from the
study of Onken et al.\ (2004). The two samples are well matched in
terms of host galaxy morphology, suggesting that morphological
selection effects are not the dominant component. However, larger
samples of AGNs with determined BH mass, stellar velocity dispersion,
and host galaxy morphology are needed both locally and at high-redshift to
improve the understanding of selection effects. Finally, if cosmic
evolution is the correct explanation, the observed offset would
support earlier growths of supermassive BHs in galaxies with mass
scales of $\langle\sigma\rangle$=170 km s$^{-1}$. This could be
evidence for `downsizing' in the BH-galaxy coevolution i.e.  more
massive galaxies arrive at the local relationship early in time.  This
scenario can be further investigated with a sample of AGN host
galaxies with a range in mass at fixed redshifts.

\acknowledgments

We thank the referee for useful suggestions.
We thank Todd Boroson for providing the I Zw 1 Fe II template; Brad
Peterson and Marianne Vestergaard for useful suggestions and for
sending us the RMS and mean spectra of several local Seyfert galaxies
with measured reverberation BH masses; Aaron Barth for numerous
stimulating conversations and for providing us with high S/N
Palomar/Hale spectra of 3 Seyfert galaxies. We acknowledge stimulating
conversations with Peng Oh, Robert (Ski) Antonucci, and Giovanni
Fossati. This work is based on data collected at Keck Observatory --
operated by Caltech and the University of California -- and with the
Hubble Space Telescope operated by AURA under contract from NASA. This
project is made possible by the wonderful public archive of the Sloan
Digital Sky Survey.  We acknowledge financial support by NASA through
HST grant GO-10216.

\clearpage

\begin{deluxetable}{lcrrrrrrrrr}
\tablewidth{0pt}
\tablecaption{Journal of observations}
\tablehead{
\colhead{Run} &
\colhead{Date} &
\colhead{Grating} &
\colhead{Slit Width}&
\colhead{Seeing}  &
\colhead{Conditions}     \\
\colhead{}       &
\colhead{}       &
\colhead{line mm$^{-1}$}       &
\colhead{arcsec}           &
\colhead{arcsec}         &  
\colhead{}         \\
\colhead{(1)} &
\colhead{(2)} &
\colhead{(3)} &
\colhead{(4)} &
\colhead{(5)} &
\colhead{(6)}}  
\tablecolumns{6}
\startdata
1 & 2003 Mar 6     &  900 &1.5 & $\sim$1        & cirrus \\
2 & 2003 Sep 3     &  900 &1.5 & $\sim$1        & cirrus \\
3 & 2004 May 14    &  900 &1   & $\sim$1        & cirrus \\
4 & 2004 May 22    &  831 &1   & $\sim$0.8      & clear \\
5 & 2005 Jul 7,8   &  900/831 &1   & 0.7-0.9 & clear  \\
\enddata
\label{T_journal}
\tablecomments{
Col. (1): Observing run.
Col. (2): Observing date.
Col. (3): Grating.  
Col. (4): Slit width.
Col. (5): Seeing FWHM.
Col. (6): Conditions.
}
\end{deluxetable}

\begin{deluxetable}{lcrrrrrrr}
\tablewidth{0pt}
\tablecaption{Targets and Exposures}
\tablehead{
\colhead{Name}        &
\colhead{z}           &
\colhead{RA (J2000)}     &
\colhead{DEC (J2000)}    &
\colhead{r'}    &
\colhead{Run}         &
\colhead{Exp.}       &
\colhead{S/N}         \\
\colhead{(1)} &
\colhead{(2)} &
\colhead{(3)} &
\colhead{(4)} &
\colhead{(5)} &
\colhead{(6)} &
\colhead{(7)} &
\colhead{(8)} }
\tablecolumns{8}
\startdata
S01& 0.3596       &  15 39 16.23 &+03 23 22.06 & 18.94 & 2,5& 7200  & 69 \\
S02& 0.3544$^{1}$ &  16 11 11.67 &+51 31 31.12 & 18.94 & 2  & 3000  & 44 \\
S03& 0.3583$^{1}$ &  17 32 03.11 &+61 17 51.96 & 18.26 & 2  & 1800  & 55 \\
S04& 0.3580       &  21 02 11.51 &-06 46 45.03 & 18.75 & 2  & 2400  & 47 \\
S05& 0.3531       &  21 04 51.85 &-07 12 09.45 & 18.43 & 2,5& 12600 & 111 \\
S06& 0.3689       &  21 20 34.19 &-06 41 22.24 & 18.53 & 2  & 3000  & 31 \\
S07& 0.3520       &  23 09 46.14 &+00 00 48.91 & 18.13 & 2,5& 7200  & 100 \\
S08& 0.3591       &  23 59 53.44 &-09 36 55.53 & 18.61 & 2  & 2400  & 54 \\
S09& 0.3548       &  00 59 16.11 &+15 38 16.08 & 18.33 & 2  & 2700  & 39 \\
S10& 0.3506$^{1}$ &  01 01 12.07 &-09 45 00.76 & 17.91 & 2  & 600   & 52 \\
S11& 0.3562       &  01 07 15.97 &-08 34 29.40 & 18.43 & 2,5& 4800  & 110 \\
S12& 0.3575       &  02 13 40.60 &+13 47 56.06 & 18.18 & 2  & 1800  & 38 \\
S21& 0.3534$^{1}$ &  11 05 56.18 &+03 12 43.26 & 17.50 & 3  & 1500  & 70 \\
S23& 0.3515       &  14 00 16.66 &-01 08 22.19 & 18.22 & 3,5& 5400  & 67 \\
S24& 0.3621       &  14 00 34.71 &+00 47 33.48 & 18.43 & 3,5& 9600  & 98 \\
S26& 0.3691       &  15 29 22.26 &+59 28 54.56 & 18.93 & 3  & 7200  & 47 \\
S27& 0.3667$^{1}$ &  15 36 51.28 &+54 14 42.71 & 18.87 & 3  & 7200  & 41 \\
S28& 0.3682       &  16 11 56.30 &+45 16 11.04 & 18.78 & 4,5& 5760  & 73 \\
S29& 0.3575$^{1}$ &  21 58 41.93 &-01 15 00.33 & 18.87 & 4  & 3600  & 56 \\
S99& 0.3690       &  16 00 02.80 &+41 30 27.00 & 18.78 & 1  & 4800  & 42 \\
\enddata
\label{T_target}
\tablecomments{
Col. (1): Target ID.  
Col. (2): Redshift from stellar absorption lines.
Col. (3): RA.
Col. (4): DEC.
Col. (5): Extinction corrected $r'$ AB magnitude from SDSS photometry.
Col. (6): Observing run.
Col. (7): Total exposure time in seconds.
Col. (8): Signal-to-noise ratio per 0.85\AA\, pixel of the combined spectrum (average in the 6900-7400\AA spectral region).}
\tablerefs{
a) redshift from SDSS DR4.
}
\end{deluxetable}

\begin{deluxetable}{lcrrrrrrr}
\tablewidth{0pt}
\tablecaption{Observed and derived properties}
\tablehead{
\colhead{Name}               &
\colhead{$\sigma_{H\beta}$} &
\colhead{$\sigma_{H\beta}$} &
\colhead{$f_{5100}$}    &
\colhead{$\lambda L_{5100}$}    &
\colhead{log M$_{\rm BH}/M_{\odot}$}  &
\colhead{$\sigma$}    &
\colhead{Type}           \\
\colhead{} &
\colhead{\AA} &
\colhead{km s$^{-1}$ } &
\colhead{$10^{-17}$ erg cm$^{-2}$ \AA$^{-1}$ s$^{-1}$} &
\colhead{$10^{44}$ erg s$^{-1}$} &
\colhead{}      &
\colhead{km s$^{-1}$}      &
\colhead{}    \\
\colhead{(1)} &
\colhead{(2)} &
\colhead{(3)} &
\colhead{(4)} &
\colhead{(5)} &
\colhead{(6)} &
\colhead{(7)} &
\colhead{(8)} }
\tablecolumns{8}
\startdata
S01  &   46.6$\pm$     0.5  &  2116. $\pm$    21.  &  5.83$\pm$     0.01  &  1.77$\pm$     0.01  &   8.20  & 132$\pm $   8  &Sb           \\
S02  &   43.7$\pm$     0.9  &  1992. $\pm$    43.  &  5.35$\pm$     0.02  &  1.57$\pm$     0.01  &   8.11  & -                &E/S0          \\
S03  &   41.2$\pm$     0.5  &  1872. $\pm$    21.  & 10.77$\pm$     0.03  &  3.25$\pm$     0.01  &   8.28  & -                &Sb           \\
S04  &   52.2$\pm$     0.6  &  2367. $\pm$    29.  &  7.83$\pm$     0.02  &  2.38$\pm$     0.01  &   8.39  & 186$\pm $   8  &Sa           \\
S05  &   67.8$\pm$     0.7  &  3090. $\pm$    33.  &  9.44$\pm$     0.00  &  2.74$\pm$     0.01  &   8.66  & 132$\pm $   5  &Sb           \\
S06  &   48.8$\pm$     0.9  &  2198. $\pm$    43.  &  9.07$\pm$     0.01  &  2.95$\pm$     0.01  &   8.39  & 169$\pm $  14  &Sb           \\
S07  &   51.4$\pm$     0.5  &  2345. $\pm$    24.  & 11.25$\pm$     0.01  &  3.23$\pm$     0.01  &   8.47  & 145$\pm $  13  &Sc           \\
S08  &   36.0$\pm$     0.6  &  1635. $\pm$    26.  &  8.72$\pm$     0.02  &  2.63$\pm$     0.01  &   8.10  & 187$\pm $  11  &Sa           \\
S09  &   43.0$\pm$     0.3  &  1960. $\pm$    16.  & 11.06$\pm$     0.02  &  3.23$\pm$     0.01  &   8.32  & 187$\pm $  15  &M            \\
S10  &   44.6$\pm$     0.4  &  2037. $\pm$    18.  & 14.62$\pm$     0.04  &  4.18$\pm$     0.01  &   8.43  & -                &Sb           \\
S11  &   40.1$\pm$     0.2  &  1822. $\pm$     7.  &  9.23$\pm$     0.01  &  2.74$\pm$     0.00  &   8.20  & 127$\pm $   9  &S0           \\
S12  &   73.7$\pm$     0.1  &  3346. $\pm$     5.  & 12.84$\pm$     0.02  &  3.87$\pm$     0.01  &   8.83  & 173$\pm $  22  &Sb           \\
S21  &   68.8$\pm$     0.5  &  3137. $\pm$    22.  & 26.54$\pm$     0.46  &  7.73$\pm$     0.13  &   8.99  & -                &M            \\
S23  &   73.3$\pm$     0.6  &  3345. $\pm$    29.  & 12.10$\pm$     0.01  &  3.46$\pm$     0.01  &   8.80  & 172$\pm $   8  &Sb           \\
S24  &   61.6$\pm$     0.4  &  2792. $\pm$    20.  &  9.96$\pm$     0.01  &  3.08$\pm$     0.01  &   8.61  & 214$\pm $  10  &Sb           \\
S26  &   41.6$\pm$     1.1  &  1871. $\pm$    50.  &  4.97$\pm$     0.08  &  1.62$\pm$     0.03  &   8.07  & 128$\pm $   8  &Sb           \\
S27  &   38.9$\pm$     0.5  &  1754. $\pm$    21.  &  6.18$\pm$     0.10  &  1.98$\pm$     0.03  &   8.07  & -                &M            \\
S28  &   54.4$\pm$     0.4  &  2450. $\pm$    17.  &  7.33$\pm$     0.01  &  2.37$\pm$     0.01  &   8.42  & 210$\pm $  10  &             \\
S29  &   45.0$\pm$     0.9  &  2044. $\pm$    41.  &  6.16$\pm$     0.01  &  1.85$\pm$     0.01  &   8.18  & -                &             \\
S99  &   70.9$\pm$     2.1  &  3196. $\pm$    94.  &  8.02$\pm$     0.01  &  2.59$\pm$     0.01  &   8.67  & 224$\pm $  12  &S0/a          \\
\enddata
\label{data}          
\tablecomments{
Col. (1): AGN name.  
Col. (2): Second moment of H$\beta$ in the observed frame.
Col. (3): Second moment of H$\beta$ in km s$^{-1}$. Following Peterson et al. (2004), FWHM can be obtained by multiplying a factor of 2.03 {$\pm$} 0.59.
Col. (4): Observed flux at 5100(1+$z$)\AA, calibrated with SDSS photometry.
Col. (5): Rest frame luminosity at 5100\AA.
Col. (6): Logarithm of BH mass in solar units. Estimated uncertainty is 0.4 dex (Vestgaard 2002).
Col. (7): Stellar velocity dispersion with uncertainty.
Col. (8): Host galaxy morphological type determined from HST imaging (Paper II). M identifies merging galaxies.
}
\end{deluxetable}

\clearpage

\clearpage
\plotone{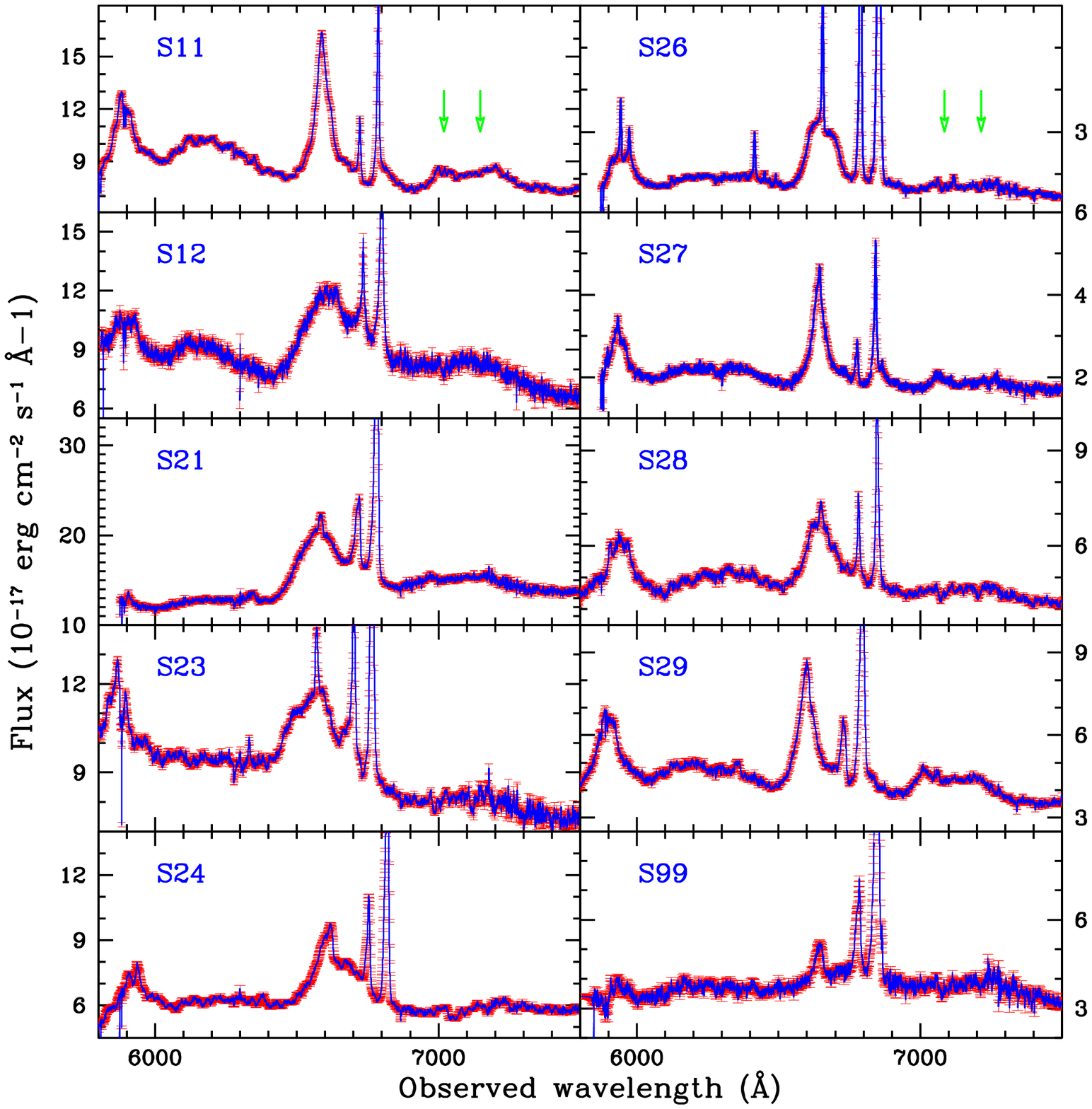}
\centerline{Fig. 1. --- Continued.}
\clearpage

\begin{figure}
\plotone{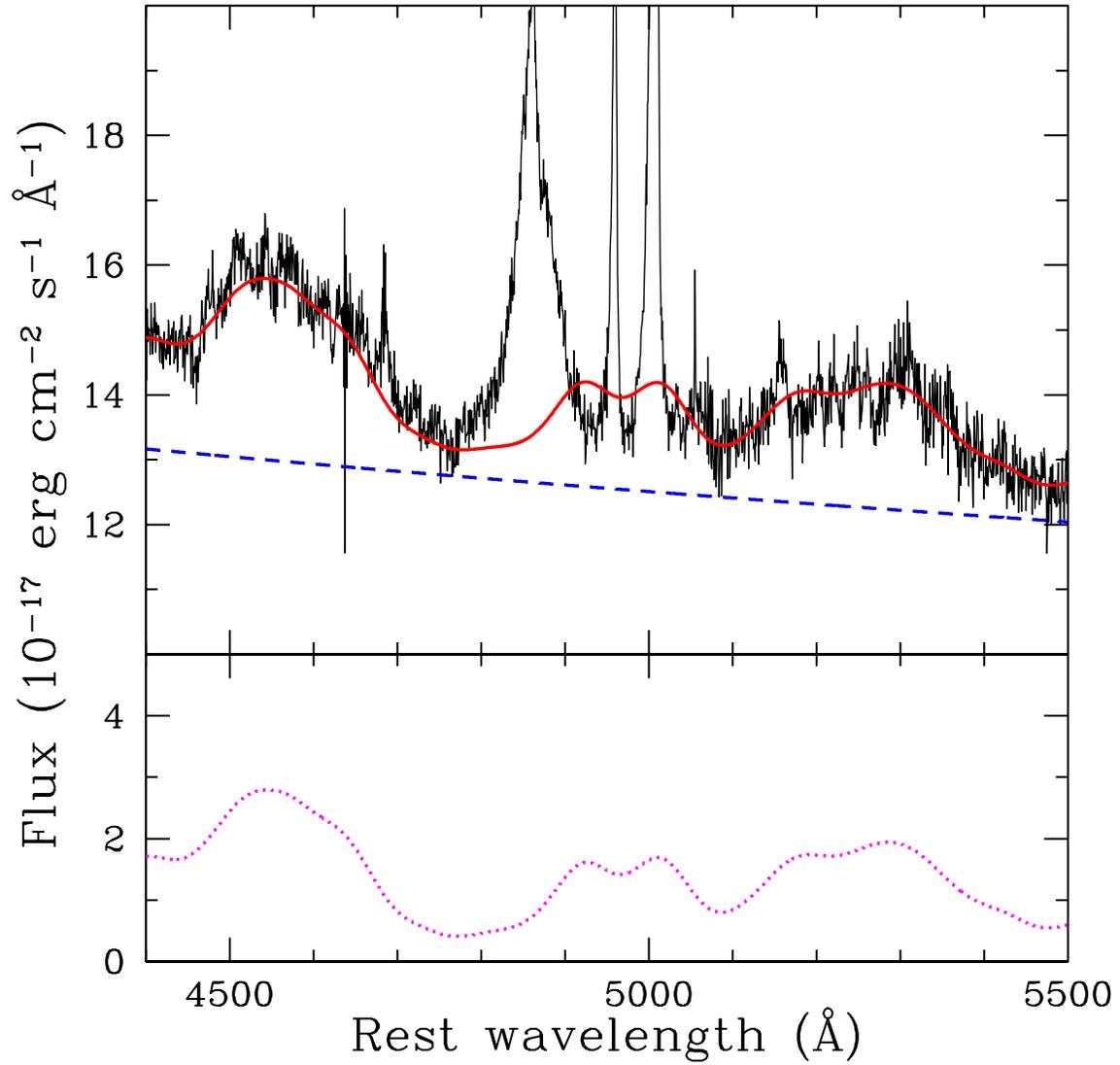}
\caption{Example of broad Fe II emission fit with I Zw 1 template and
a featureless continuum.  The best fit model (solid {\it red} line;
top panel), composed of a broadened Fe II template (dotted {\it
magenta} line; lower panel) and AGN+stellar continuum (dashed {\it
blue}; upper panel), is compared with the observed spectrum ({\it
black} histogram; upper panel). Three free parameters, i.e. continuum
slope, Fe II width, and Fe II strength are determined by minimizing the
$\chi^{2}$ in the spectral region 5100-5500 \AA.}
\end{figure}

\begin{figure*}
\plotone{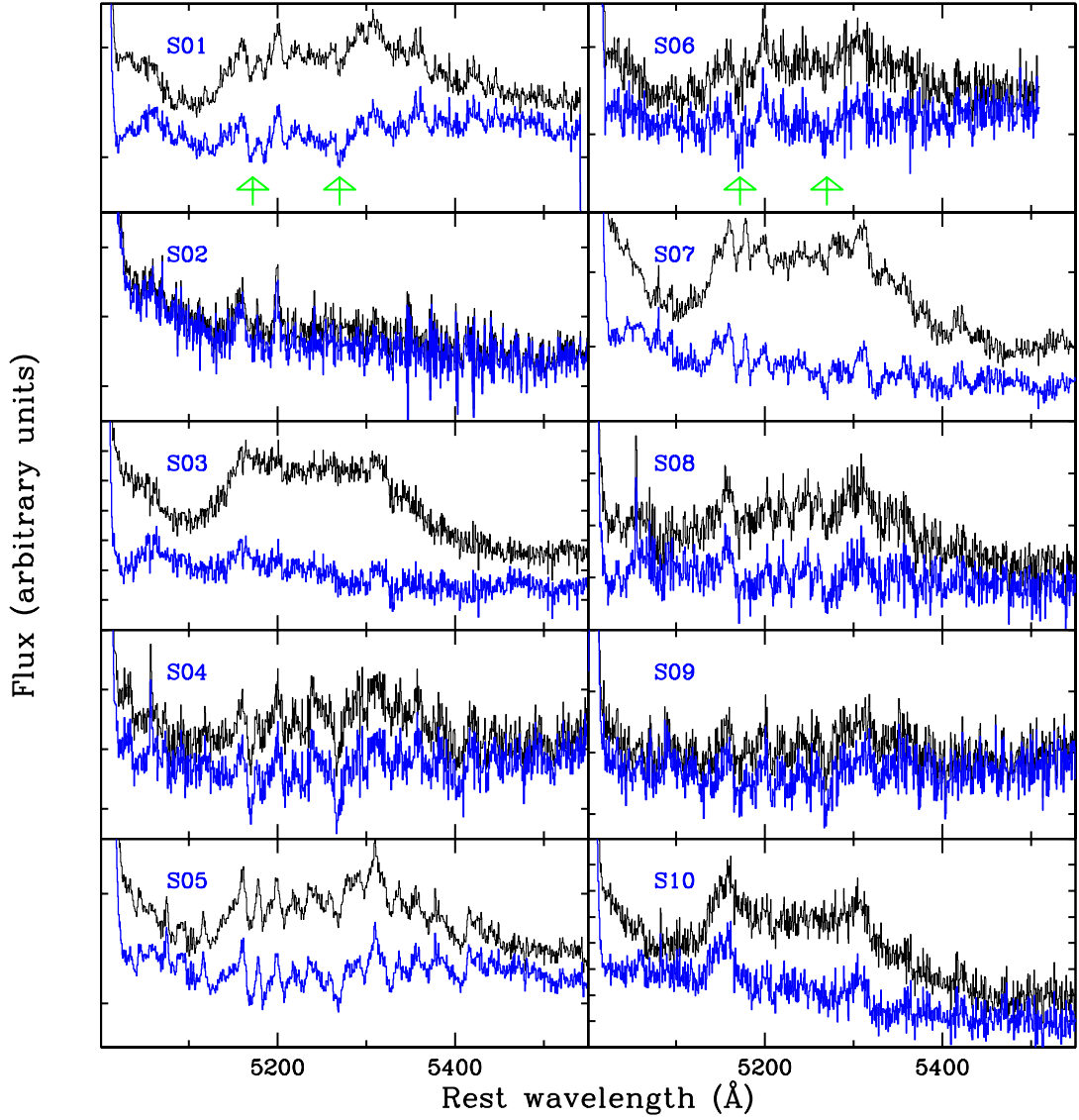}
\caption{Rest frame spectra before (upper black histogram) and after
(lower blue histogram) Fe II subtraction. Note the large range of Fe II
emission strength. 
Green arrows indicate the
location of the stellar features, Mgb (5175 \AA) and Fe (5270 \AA).\label{fig_Fe IIsub}}
\end{figure*}

\clearpage
\plotone{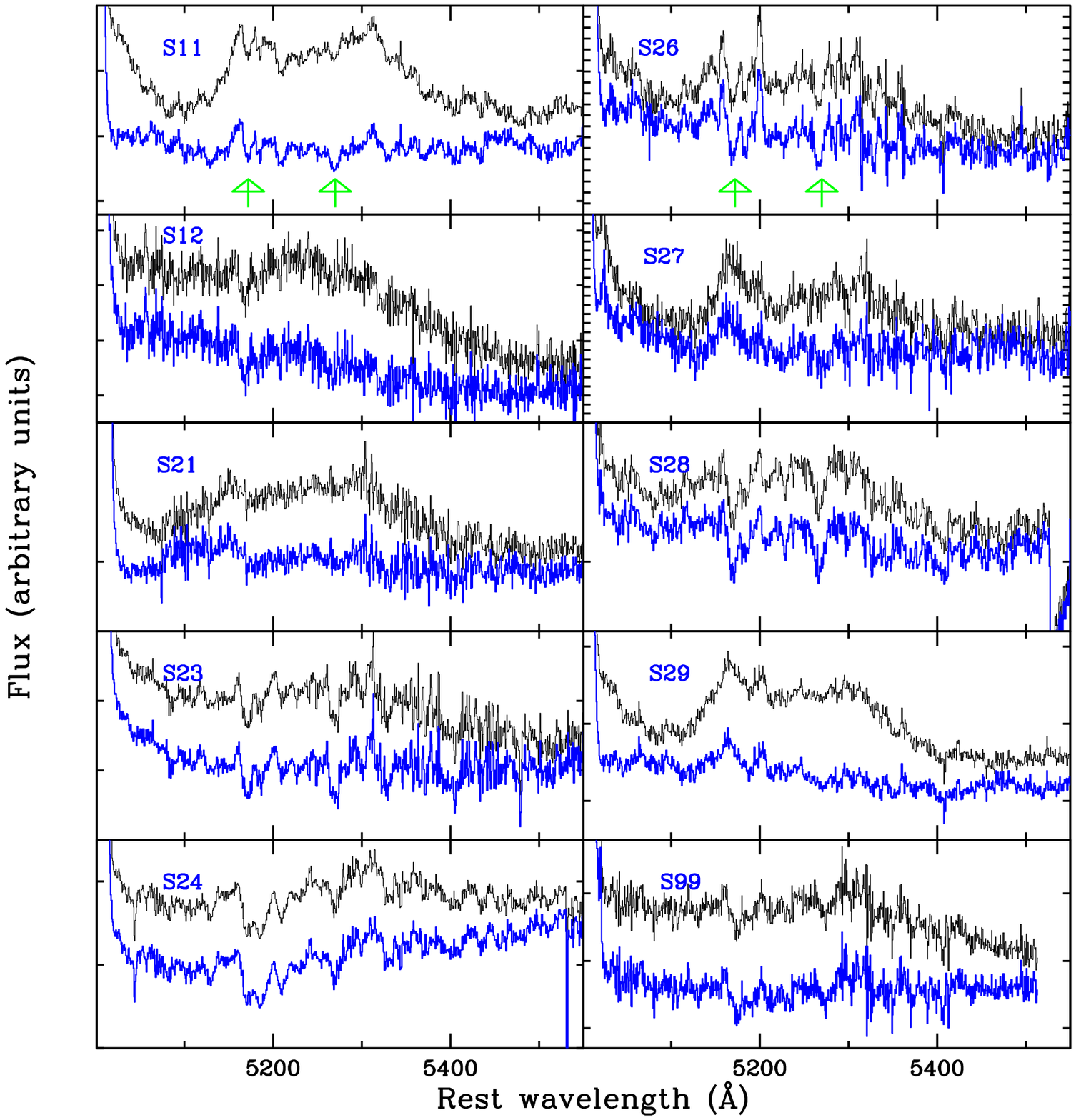}
\centerline{Fig. 3. --- Continued.}
\clearpage

\begin{figure}
\plotone{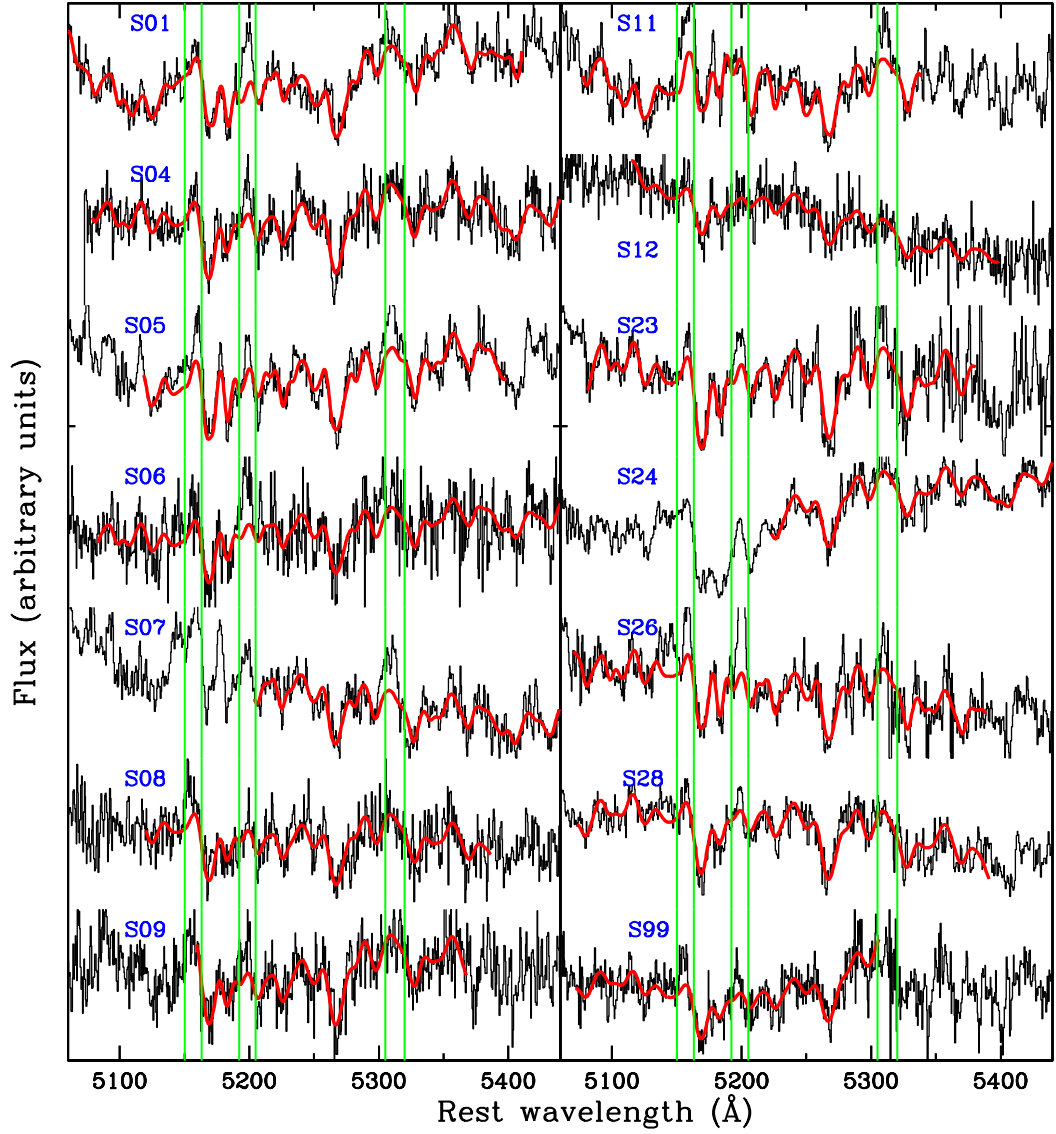}
\caption{Velocity dispersion measurements. The region including the
main stellar features around Mgb (5175)and Fe (5270) is shown (black histogram)
together with the best fit template (red thick line). The regions around narrow
AGN emission lines -- identified by green vertical lines -- are masked
out before fitting.}
\label{fig_dis}       
\end{figure}

\begin{figure}
\plotone{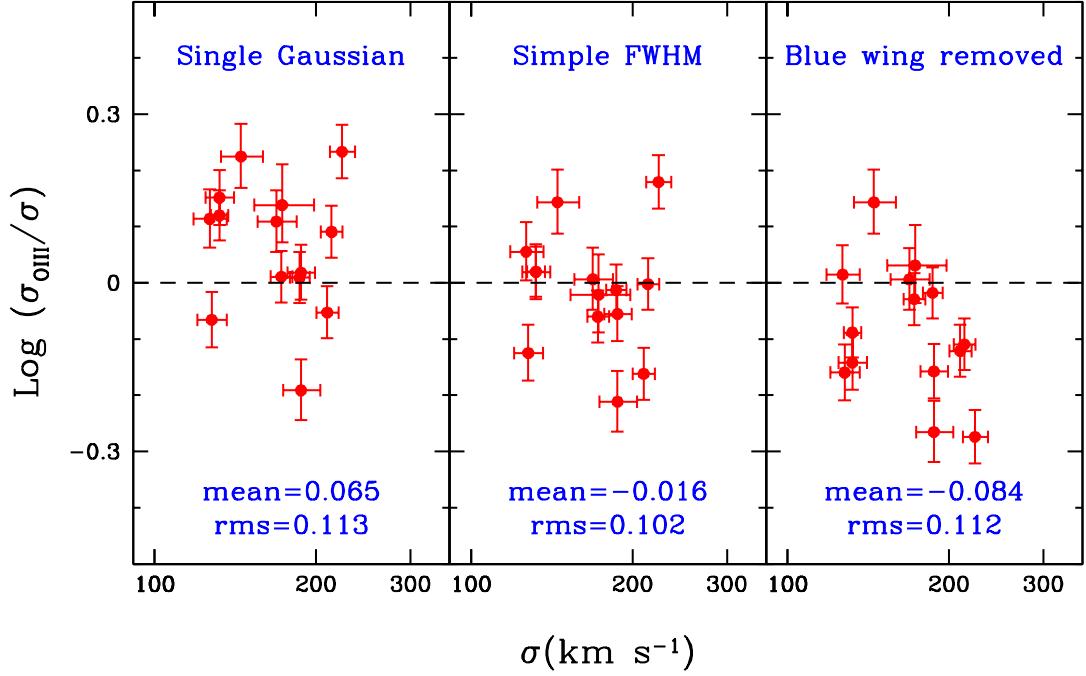}
\caption{Comparison of velocity dispersion measured from stellar
features and from the narrow [\ion{O}{3}] line. From left to right, we
compare with the stellar velocity dispersion a single gaussian fit to
the emission line, the FWHM divided by 2.35, and a double gaussian
fit. Note that the single gaussian fit provides a very bad fit to the
data because the narrow lines are asymmetric with a prominent blue
wing.}
\label{fig_OIII_sig}
\end{figure}

\begin{figure}
\plotone{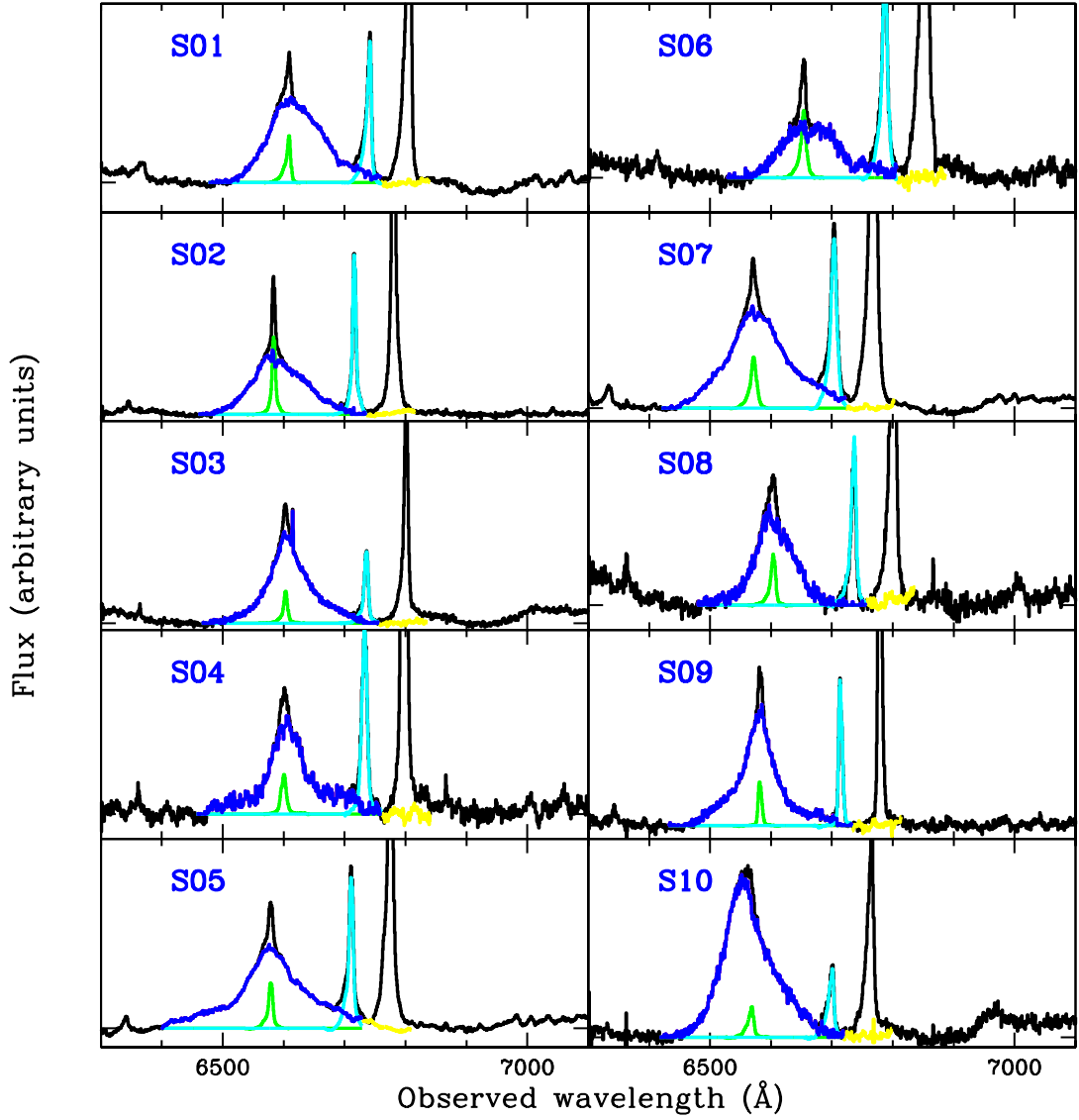}
\caption{Determination of the second moment of broad H$\beta$. After
removing the continuum (cyan horizontal line), [\ion{O}{3}]$\lambda$5007
is rescaled and blueshifted to subtract [\ion{O}{3}]$\lambda$4959 (cyan
line) and the narrow core of H$\beta$ (green line). The second moment
is measured on the residual broad H$\beta$ (blue histogram). See
Section~\ref{ssec:hbeta} for details.}
\label{fig_hbeta1}       
\end{figure}

\clearpage
\plotone{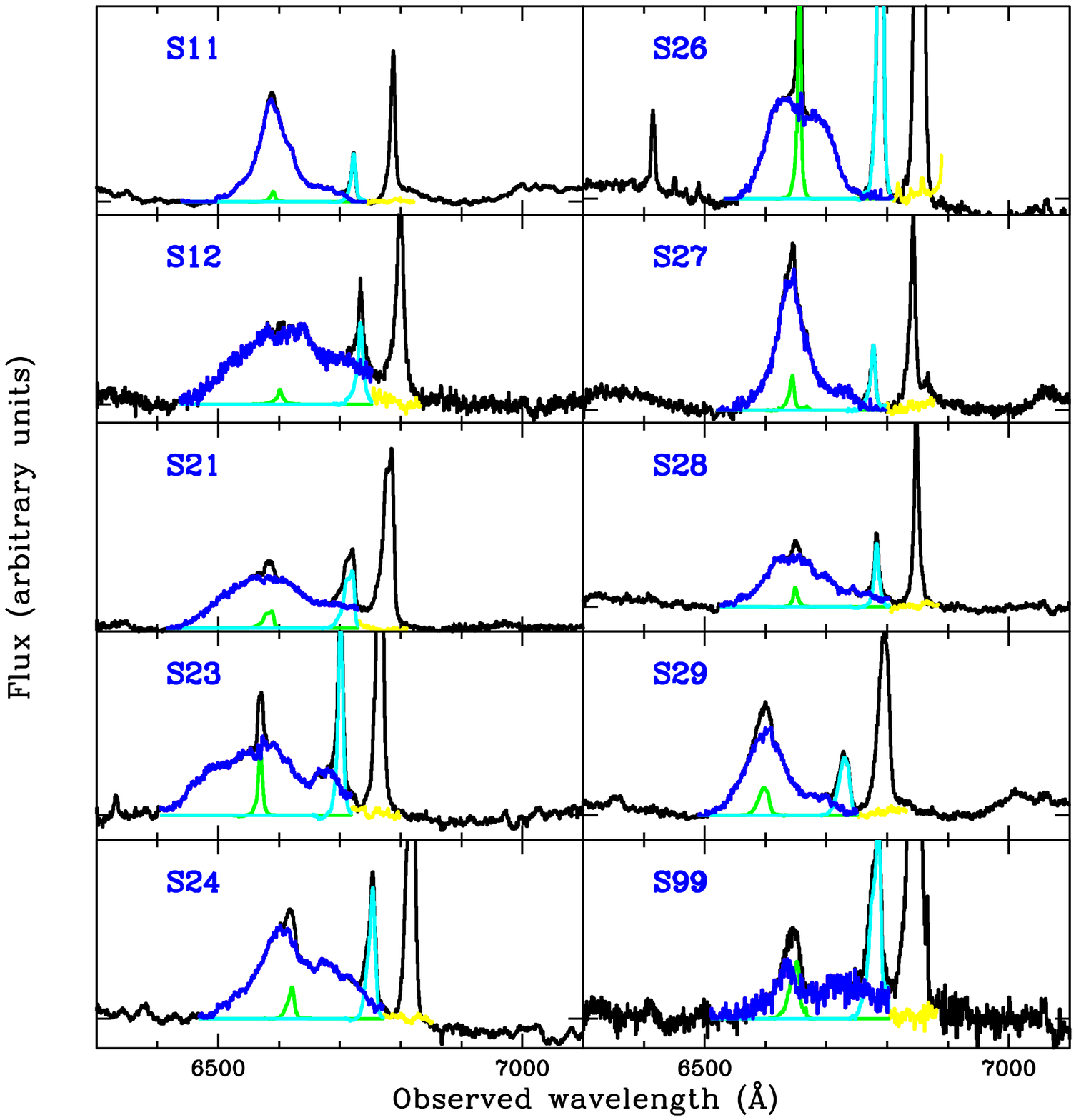}
\center{Fig. 6. --- Continued.}
\clearpage

\begin{figure}
\plotone{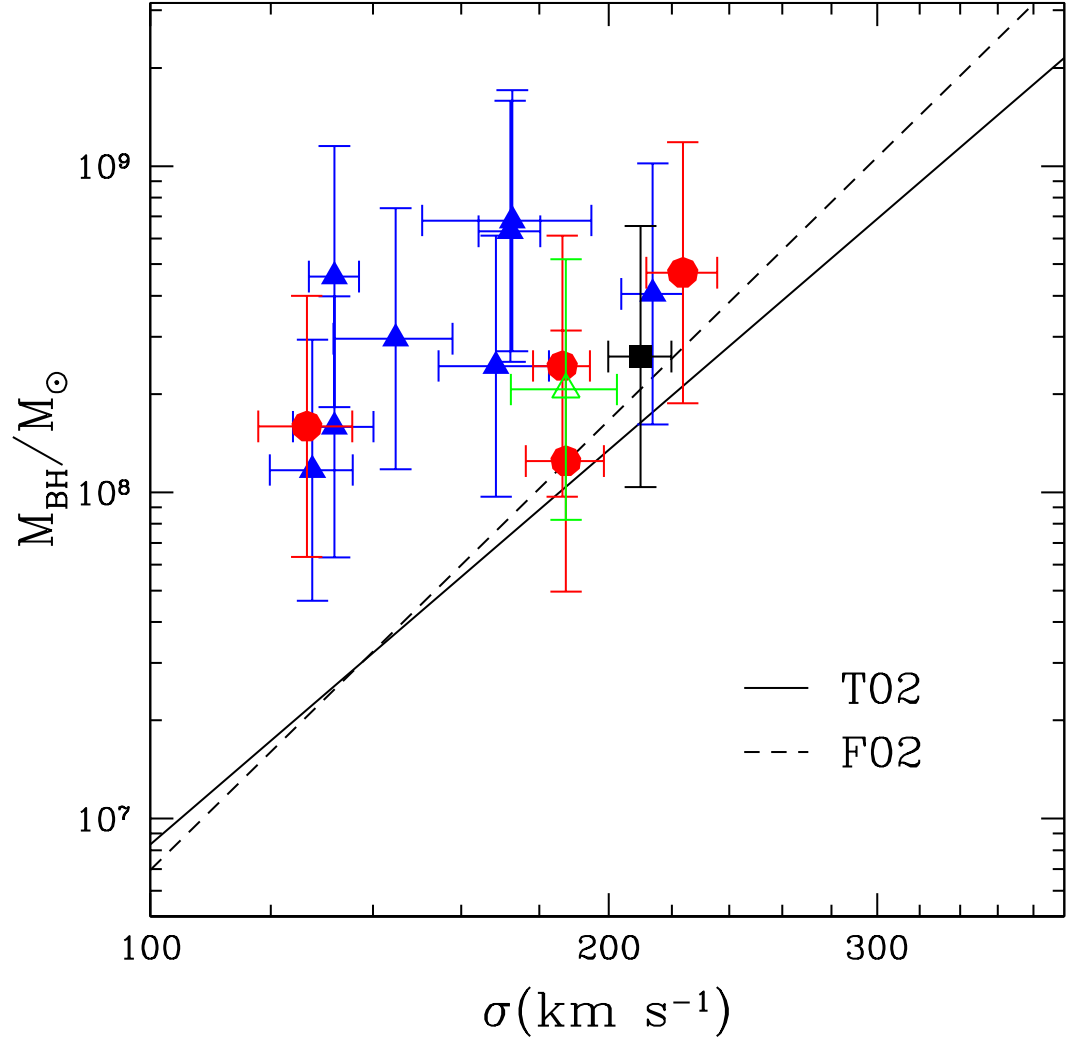}

\caption{The M$_{\rm BH}-\sigma$ relation for our sample of Seyfert
galaxies at $z=0.36$. The local relationships for quiescent galaxies
as measured by Tremaine et al. (2002; T02; solid line) and by
Ferrarese (2002; F02; dashed line) are shown for comparison. Note that
the points at $z=0.36$ lie above the local relationship, consistent
with smaller velocity dispersions for given BH mass.
Host galaxy morphological types from
HST imaging are also shown: red circles identify early-type galaxies
(E, S0, and Sa); blue solid triangles late-type galaxies; a green open
triangle highly disturbed merging systems; a black solid square
indicates no HST image available.}
\label{fig_msigma}
\end{figure}

\begin{figure}
\plotone{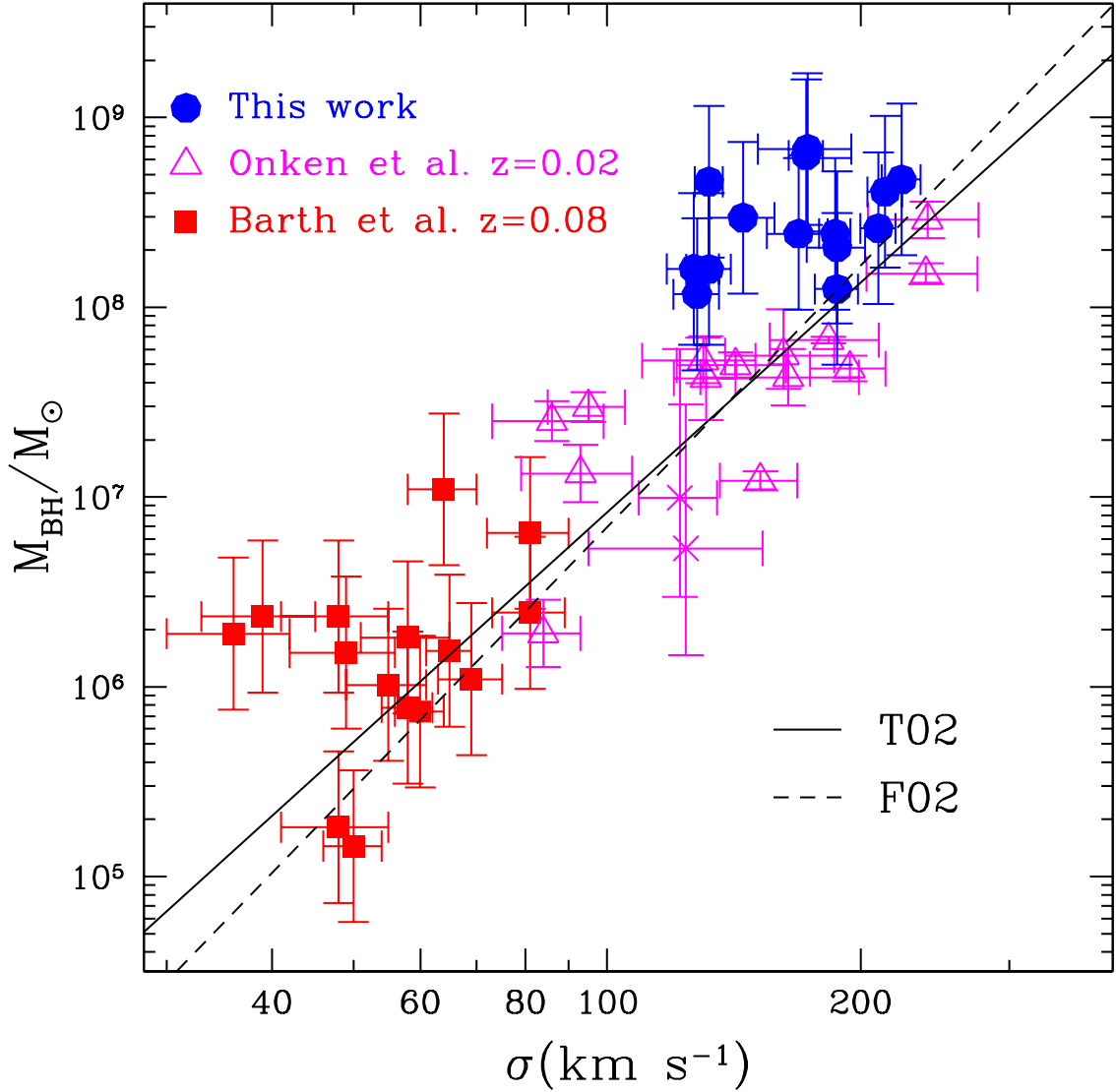}
\caption{The M$_{\rm BH}-\sigma$ relation of active galaxies. The
symbols represent 14 Seyferts at $z=0.36$ from this work (blue
circles), 15 dwarf Seyfert galaxies at $z \sim 0.08$ from Barth et
al.\ (2005; red squares), 14 local AGNs with BH masses measured via
reverberation mapping from Onken et al. (2004; magenta triangles; two
additional objects, excluded by Onken et al. and for consistency in
our work, are shown as crosses). The local relationships of quiescent
galaxies are shown for comparison as a solid (Tremaine et al. 2002)
and dashed (Ferrarese 2002) line.  Note that BH masses from this work
and the Barth et al. sample adopt the same calibration of the shape
factor as calculated by Onken et al. and therefore the relative
position of the three samples along the y-axis is independent of the
shape factor, provided it is redshift independent. Also, the shape
factor given by Onken et al.\ 2004 is derived by requiring the local
relationship for active and quiescent galaxies to be the same (see
Onken et al.\ 2004 for details).  }
\label{fig_all}
\end{figure}

\begin{figure}
\epsscale{.9}\plotone{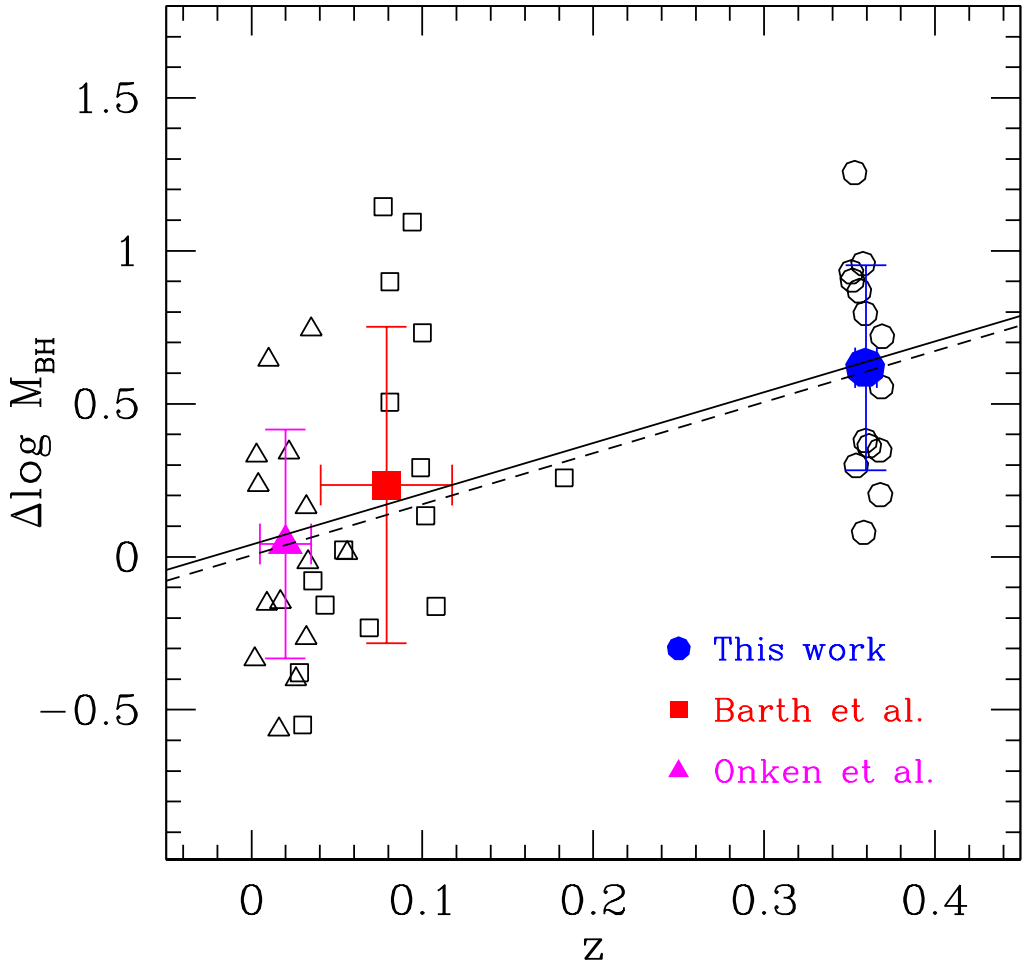}\epsscale{1}
\caption{Offset from the local M$_{\rm BH}-\sigma$ relation of
Tremaine et al.\ (2002) for three sets of data (Onken et al. 2004 at
$z \sim 0.02$; Barth et al 2005 at $z \sim 0.08$; our sample at
$z=0.36$). Large solid points with error bars represent the average
and rms scatter for the three samples. The best linear fit to the data
are shown as a solid line (for the three samples) and a dashed line
(excluding the sample of dwarfs from Barth et al. 2005). The best fit
linear relationship is $\Delta \log M_{\rm BH} = (1.66\pm0.43)z +
(0.04\pm0.09)$. The rms scatter of the $z=0.36$ sample is 0.35 dex,
similar to that of the Onken et al. points and to the estimated
uncertainty on the BH mass determination via ECPI. The average offset
of the $z=0.36$ points is $0.62\pm0.10$ dex in BH mass corresponding
to 0.15 dex in $\Delta \log \sigma$.  Adopting the Ferrarese (2002)
relationship leaves virtually unchanged the offset for our points
($0.57\pm0.11$) and for the Onken et al. points, while increasing the
offset of the Barth et al. (2005) sample significantly due to the
large difference of the relationships for BH masses of order $10^{6}
M_{\odot}$ (see Figure 8). Including only our points and the Onken et
al. points the offset with respect to the Ferrarese (2002) relation is
$\Delta \log M_{\rm BH}=(1.55\pm0.46)z+0.01\pm0.12$.}
\label{fig_evolution}
\end{figure}

\begin{figure}
\plotone{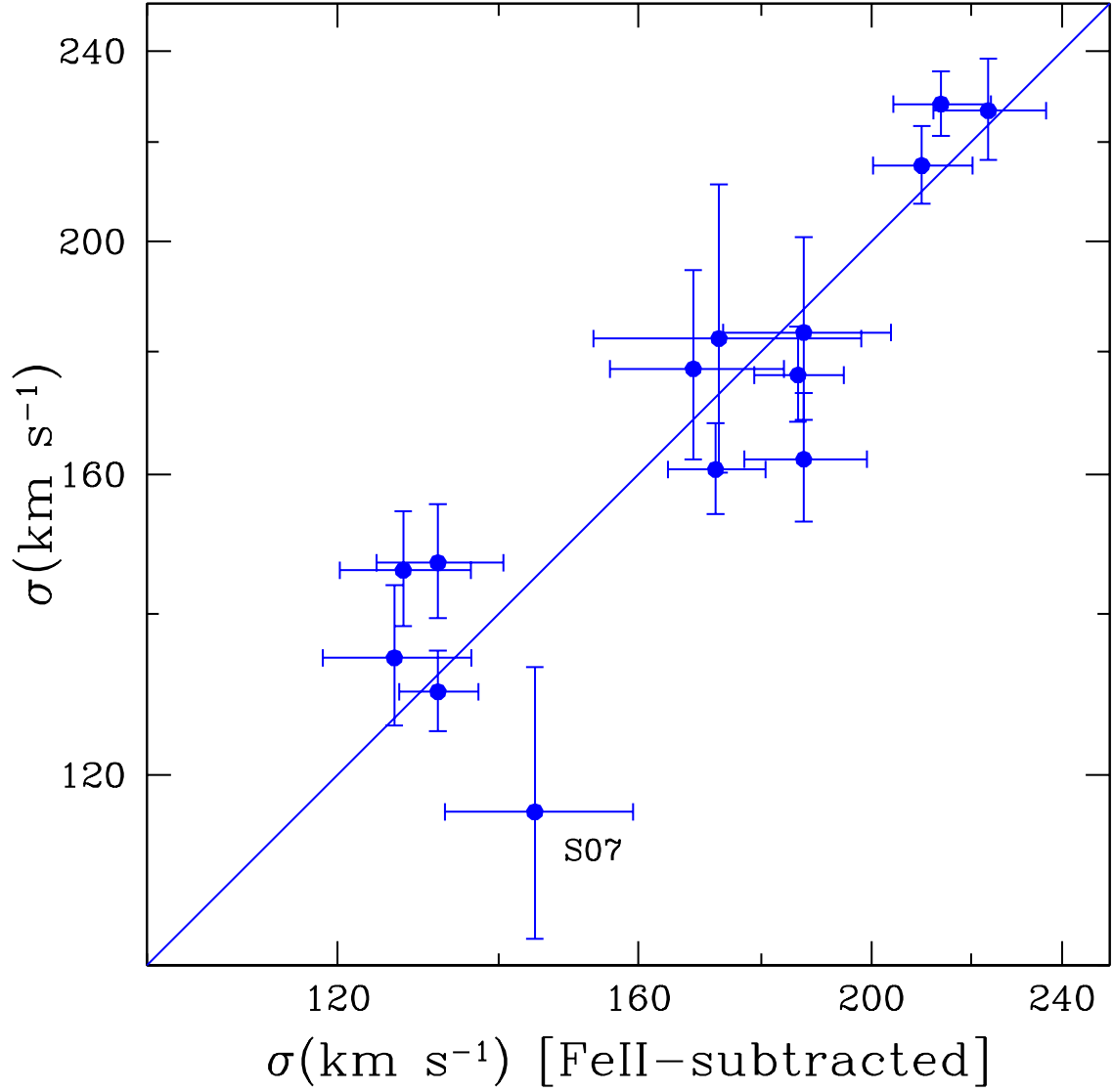}
\caption{Comparison of stellar velocity dispersion measurements with
and without the broad Fe II emission subtraction. The two measurements are
consistent with mean offset of 0.01 dex and a rms scatter 0.04 dex.
The most deviant point, S07, shows the strongest Fe II
emission, consistent with showing the largest difference ($\sim$0.1 dex).  }
\label{fig_com_sig}
\end{figure}

\begin{figure}
\plotone{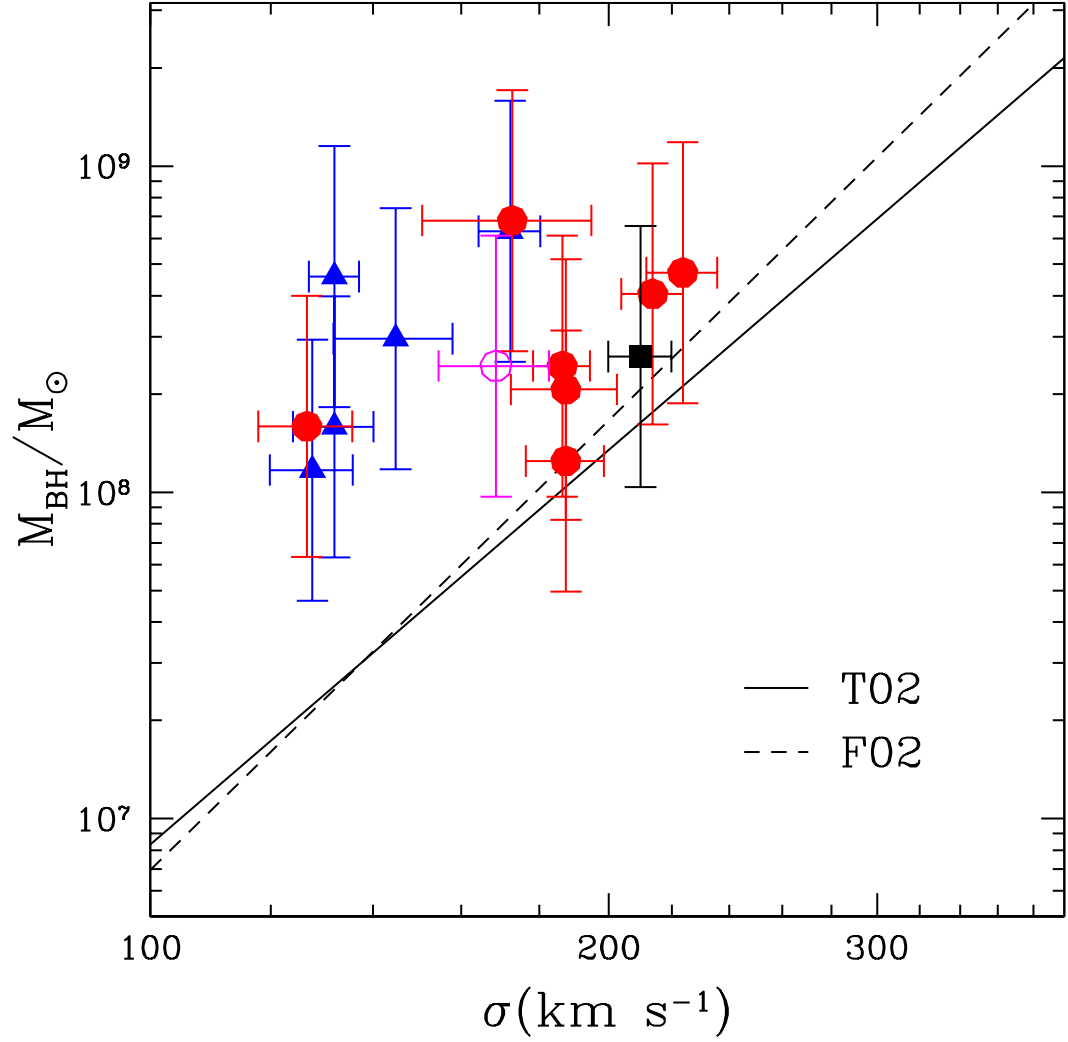}
\caption{The M$_{\rm BH}-\sigma$ relation depending on host galaxy
inclination, as determined from HST images. Face-on galaxies (blue
triangles) show smaller velocity dispersions compared to edge-on
galaxies (open circle) and intermediate inclination + early-type galaxies 
(solid circles). 
A black square indicates an object without HST image.
The offset without the face-on galaxies is $0.45 \pm 0.10$ in BH mass, 
0.17 dex smaller than that of the total sample.
The offset is found to be $0.92 \pm 0.10$ and $0.43 \pm 0.10$,
respectively, for face-on galaxies and intermediate / undefined galaxies.
}
\label{fig_msigma_inc}
\end{figure}

\begin{figure}
\plotone{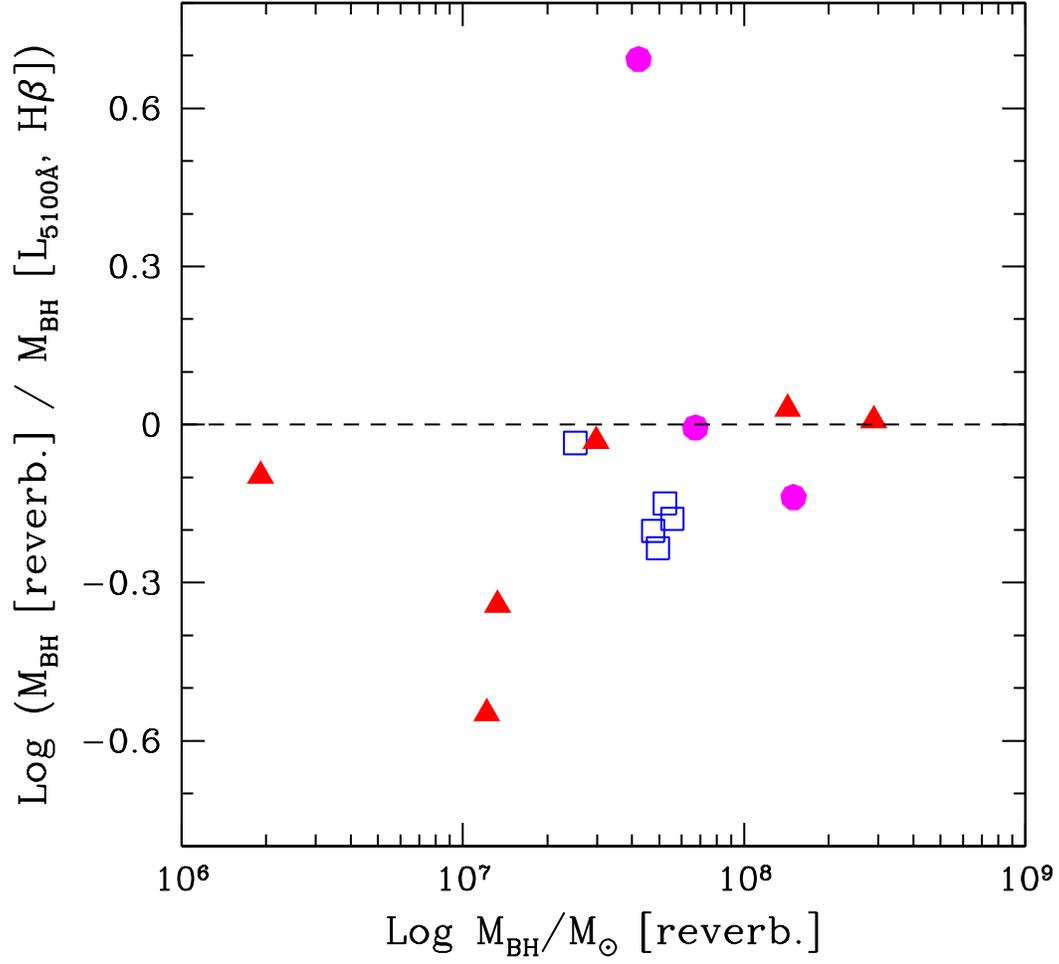}
\caption{ Comparison of BH mass obtained from reverberation results
(from Peterson et al. 2004) and the ECPI method 
(i.e. from H$\beta$ width and $L_{5100}$ using Eq. 2)
using single-epoch spectra (circles: spectra provided by A. Barth; 
triangles: spectra from International AGN Watch website)
or mean spectra (squares: provided by B. Peterson).
The mean offset is $0.09 \pm 0.07$ dex,
indicating our H$\beta$ width measurements from
single-epoch data are consistent with those from the rms spectra.
}
\label{fig_com_rev}
\end{figure}

\begin{figure}
\plotone{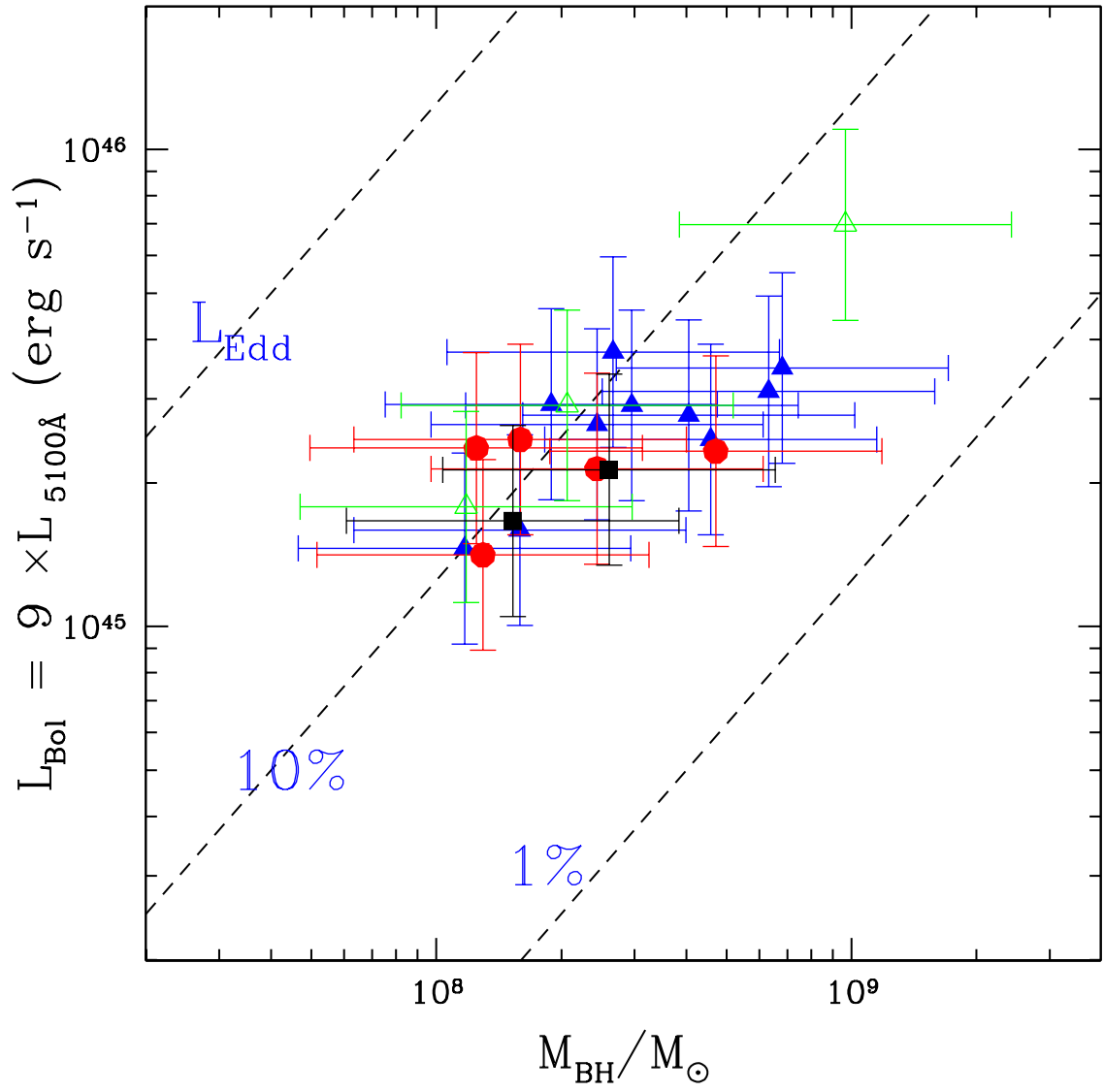}
\caption{ Bolometric luminosity vs. BH mass.  Bolometric
luminosities are calculated from optical luminosity multiplied by 9.
Eddington ratios are indicated by dashed lines.  Most of the observed
Seyfert galaxies show low Eddington ratios, implying a low accretion
rate. Symbols are same as in Fig.~7.}
\label{fig_LEDD}
\end{figure}

\end{document}